\documentclass[preprint,12pt]{elsarticle}

\usepackage{amssymb}
\usepackage[hidelinks]{hyperref}
\usepackage{amsmath,amssymb,amsfonts}
\usepackage[inline]{enumitem}
\usepackage{multirow}
\usepackage{blindtext}
\usepackage[utf8]{inputenc}
\usepackage{graphbox} 
\usepackage{textcomp}
\usepackage[table]{xcolor}
\usepackage{algorithmic}

\usepackage{tikz} 
\usetikzlibrary{arrows.meta,positioning} 
\usepackage{verbatim} 
\usepackage{mathtools} 
\usepackage{longtable} 
\usepackage{array,graphicx} 
\usepackage{booktabs} 
\usepackage{pifont} 
\newcommand{\algorithmicbreak}{\textbf{break}} 
\newcommand{\BREAK}{\STATE \algorithmicbreak}
\usepackage{arydshln}
\usepackage{todonotes}
\usepackage[linesnumbered,ruled,vlined]{algorithm2e}

\newenvironment{enuminline}
{\begin{enumerate*}[label={\textbf{(\textit{\roman*})}}]}
{\end{enumerate*}}

\newcommand{\mdp}{\mathcal{M}}
\newcommand{\St}{\ensuremath{S}}
\newcommand{\Act}{\mathit{Act}}
\newcommand{\PM}{\mathbf{P}}
\newcommand{\policy}{\pi}
\newcommand{\GS}{\mathit{B}}
\newcommand{\Prmax}{\mathit{Pr}^{\max}}
\newcommand{\reach}{\mathit{reach}}
\newcommand{\SDP}{\ensuremath{\mathit{SDP}}}
\newcommand{\fdd}{f_{\mathit{dd}}}
\newcommand{\powerset}{\raisebox{.15\baselineskip}{\ensuremath{\wp}}}
\newcommand{\bigO}[1]{O\left({#1}\right)}
\DeclareMathOperator*{\argmax}{argmax}

\def\RU{{RUCoP}\xspace}
\def\IRU{{L-RUCoP}\xspace}
\def\CGRU{{CGR-UCoP}\xspace}

\newcommand{\setSt}{\mathcal{S}}
\newcommand{\setCN}{\mathcal{C}} 
\newcommand{\setPaths}{\mathcal{P}}
\newcommand{\setRules}{\mathcal{R}}
\newcommand{\pred}{\mathit{pred}}
\newcommand{\pth}{\mathit{path}}
\newcommand{\cp}{\mathit{cp}}
\newcommand{\Tr}{\mathit{Tr}}
\newcommand{\LTr}{\mathit{LTr}}
\newcommand{\Prob}{\mathit{Pr}}

\newcommand{\ts}{\mathit{ts}}
\newcommand{\rc}{\mathit{rc}}
\newcommand{\Rl}{\mathit{Rl}}
\newcommand{\route}{\mathit{r}}
\newcommand{\SDPCGR}{\ensuremath{\SDP_{\mathit{CGR}}}}

\definecolor{lightgray}{RGB}{196,196,196}

\newcommand{\minisection}[1]{\medskip\par\noindent\textbf{#1}}

\newcommand{\prd}[1]{{\textcolor{red}{PRD: #1 :PRD}}}

\newcommand{\new}[1]{{\textcolor{black}{#1}}}

\begin{document}
\begin{frontmatter}

\title{Routing in Delay-Tolerant Networks under Uncertain Contact Plans}

\author[inst1]{Fernando~D.~Raverta}

\affiliation[inst1]{organization={CONICET-UNC, Argentina.},
            city={C\'ordoba},
            postcode={5000}, 
            country={Argentina}}

\author[inst1,inst2]{Juan~A.~Fraire}

\affiliation[inst2]{organization={Department of Computer Science, Saarland University},
            city={Saarbr\"ucken},
            postcode={66123}, 
            country={Germany}}
            
\author[inst1]{Pablo G. Madoery}
\author[inst1]{Ramiro A. Demasi}
\author[inst1]{Jorge M. Finochietto}
\author[inst1,inst2]{Pedro R. D'Argenio}

\begin{abstract}
Delay-Tolerant Networks (DTN) enable store-carry-and-forward data transmission in networks challenged by frequent disruptions and high latency.
Existing classification distinguishes between scheduled and probabilistic DTNs, for which specific routing solutions have been developed.
In this paper, we uncover a gap in-between where uncertain contact plans can be exploited to enhance data delivery in many practical scenarios described by probabilistic schedules available \textit{a priori}.
Routing under uncertain contact plans (\RU) is next formulated as a multiple-copy Markov Decision Process and then exported to local-knowledge (\IRU) and Contact Graph Routing extensions (\CGRU) which can be implemented in the existing DTN protocol stack.
\RU and its derivations are evaluated in a first extensive simulation benchmark for DTNs under uncertain contact plans comprising both random and realistic scenarios.
Results confirm that \RU and \IRU closely approach the ideal delivery ratio of an oracle, while \CGRU improves state-of-the-art DTN routing schemes delivery ratio up to 25\%.
\end{abstract}


\end{frontmatter}

\section{Introduction}
\label{sec:intro}

The term Delay tolerant networking (DTN) was introduced by K. Fall in 2003 to designate time-evolving networks lacking of a continuous and instantaneous end-to-end connectivity~\cite{Fall2003,RFC4838}.
Since then, DTNs have drawn much attention from many researchers due to its applicability in very distinct domains including deep space~\cite{Burleigh2003} and near Earth communication networks~\cite{Caini2011}, airborne networks~\cite{gupta2015survey}, vehicular ad-hoc  networks~\cite{BENAMAR2014141}, mobile social networks~\cite{7876231}, Internet of things~\cite{7921980} and underwater networks~\cite{Partan2007}. 
Indeed, delay and disruption conditions can be generated by long signal propagation time, regular node occlusion, high node mobility and reduced communication range and resources.

Although from diverse origins, partitions and delay in DTNs are tackled by a \textit{bundle layer} that sits above specific layers of each network family~\cite{RFC5050}.
The key feature of the bundle layer is a persistent storage on each DTN node to store-carry-and-forward \textit{bundles of data} (or simply \textit{bundles} as per DTN terminology) as transmission opportunities become available.
Since data can propagate or rest in intermediate nodes for arbitrary amounts of time, DTN protocols and applications assume no immediate response from the receiver and tend to minimize end-to-end exchanges~\cite{pottner2011performance}.
The time-evolving and partitioned nature of DTNs favor the representation of connectivity by means of \textit{contacts}, a contact being an episode of time when a node is able to transfer data to another node.

\minisection{Taxonomy}
The literature~\cite{RFC4838} classifies contacts in DTNs as:
\begin{itemize}
    \item \textit{Scheduled:}
    Contacts can be accurately predicted.
    Expected contacts can be imprinted in a \textit{contact plan} comprising an exhaustive expression of the future network connectivity~\cite{Fraire2015}.
    Such knowledge can be exploited to optimize resource utilization~\cite{Fraire2016Traffic,Fraire2015Routing,Fraire2014Fair}, medium access decisions~\cite{carosino2018integrating} and routing calculations such as in Contact Graph Routing (CGR) algorithm~\cite{FRAIRE2021102884,Araniti2015}.
    \item \textit{Probabilistic:}
    Contact patterns are dynamically inferred as network evolves in time.
    Routing is based on a topology model composed of probabilistic metrics accounting for the likelihood of meeting a given neighbour in the future~\cite{grasic2011evolution,burgess2006maxprop, jain2004routing}.
    In order to enhance delivery probability, multiple copies are sent through different paths, an approach that has also been considered for scheduled DTNs to forego the need of processing large contact plans~\cite{Feldmann2017}.
    \item \textit{Opportunistic:}
    No assumptions can be made on future contacts.
    Trivial flooding-based schemes have been used for opportunistic DTNs~\cite{Vahdat00epidemicrouting}, as well as controlled flooding such as Spray-and-Wait (S\&W) to reduce replication overhead~\cite{Spyropoulos05sprayandwait,spyropoulos2007spray}, \new{among others opportunistic path models~\cite{8737620}}.
    Also, previous research has extended scheduled routing approaches to cope with unpredictable opportunistic contacts~\cite{Burleigh2016}.
\end{itemize}

In this paper, we claim the existence of DTN under \textit{uncertain schedules} or \textit{uncertain contact plans}, which are not properly covered by the existing DTN classification:

\begin{itemize}
    \item \textit{Uncertain:} Contacts whose materialization can differ from the original plan with a given probability available \textit{a priori}.
    For example, expected contacts have a chance of being affected by well-known failure modes or by an incomplete or inaccurate (but bounded) knowledge of the system status by the time the schedule was computed.
    In other words, while in probabilistic DTNs the probability is assigned to a next-hop node (i.e, the probability of meeting a given node, based on contact history), uncertain DTNs under uncertain contact plans assign probabilities to forthcoming contacts (i.e., the probability of meeting a given node in a given time episode in the future).
\end{itemize}

\minisection{Uncertain DTNs.}
Uncertain DTNs differ from perfectly scheduled DTNs in the nature of their contacts, which are no longer certain to occur (uncertain contacts have an associated probability of existing or failing).
They also differ from probabilistic DTNs in the features of the model used to represent and reason about the network dynamics.
Instead of relying on abstract node's visibility patterns (learned on the fly), uncertain DTNs exploit time-dependant probabilistic information of the forthcoming connectivity episodes encoded in the so-called uncertain contact plan (computed in advance).
An uncertain contact plan is a probabilistic schedule that includes information regarding the probability of future contacts to diverge from the plan.
\new{The advantage of accounting for this knowledge in uncertain DTNs is that} it can be used to make specific routing, forwarding and bundle replication decisions over the most reliable routes towards a destination, thus optimizing the data delivery chances.

The different nature of probabilistic and uncertain DTNs can also be appreciated in the route structure.
Routes in probabilistic DTNs are expressed as a \textit{sequence of nodes} through which the bundle shall be forwarded.
There is no specific information on when the route hops will actually happen, just a time-averaged expectation based on inter-nodes visibility patterns. 
On the other hand, uncertain contact plans bring the notion of uncertain contact, which is also probabilistic, but encoding timing information is unavailable in traditional probabilistic schemes.
Thus, and similarly to scheduled DTNs, routes in uncertain DTNs are constructed as a \textit{sequence of uncertain contacts}, which renders a delivery probability through each path, and thus, more granular and accurate (but also challenging) decision making opportunities.

\new{Applications for uncertain DTNs include} DTN networks based on a schedule of fault-prone nodes (unreliable space networks~\cite{Fraire2017-Hindawi}), uncertain mobility patterns (public vehicle networks~\cite{kalaputapu1995modeling}), interference-sensitive communication links (cognitive radio~\cite{sahai2006fundamental}), or third-party carriers with limited availability (backbone links with known reliability~\cite{hwang1981system}).
\new{Indeed, the uncertain contact plan including contacts probabilities can be computed by} specific network models (i.e., fault-prone satellite trajectories), empirically estimated in a controlled environment (i.e., lab or simulation setup), or made available from existing statistics (i.e., interference reports).
As a result, an uncertain contact plan can be conveniently pre-computed instead of dynamically learned by nodes as in probabilistic DTNs, removing the burden of a training phase, and benefiting from highly accurate routing schemes for uncertain DTNs as introduced in this paper.

\minisection{Previous Works.}
Previous works have addressed the survivability properties of time-varying networks~\cite{Liang2017}, as well as the problem of reliable topology design in DTN~\cite{Li2015}.
However, to the best of the authors' knowledge, the problem of reliable route determination based on uncertain contact plans has been overlooked.
Authors have already studied how schedule-aware (i.e., CGR) and schedule-agnostic (i.e., S\&W) routing schemes behave under uncertain contact plans in~\cite{Fraire2017-Hindawi,Madoery2017rpic,Madoery2018} (probabilistic routings such as MaxProp~\cite{burgess2006maxprop} and Prophet~\cite{grasic2011evolution} were disregarded as they are based on learning phases during network operations).
These papers essentially showed that existing routing schemes only perform well on their respective domains (perfectly scheduled or fully opportunistic), while significant room for improvement was identified for scenarios with uncertain schedules.
In order to evaluate the potential improvement, the authors in~\cite{Raverta2018} have approached the problem with a first theoretical formulation based on probabilistic model checking techniques~\cite{BiancoA95,BaierK08,BaierAFK18}, where the contact plan with its respective fault probabilities is modelled as a Markov Decision Process (MDP).
Although this first approach provided a compelling optimal solution for single-copy routing, replication-based heuristics remained an open topic.
Exception to this statement is a recent publication that addressed the multi-copy DTN routing problem by means of approximated simulations techniques based on distributed schedulers~\cite{d2020sampling}. However, simulation techniques lack the required optimality guarantee that formal MDP models can provide.

\minisection{Contributions.}
In this paper, we present Routing under Uncertain Contact Plans (\RU), a comprehensive framework to execute reliable routing under uncertain contact plans. 
\RU embraces single copy~\cite{Raverta2018} and extends it to multiple-copy routing in an overcoming MDP model expression.
As the fact of considering multiple copies renders the focus of~\cite{Raverta2018} unsuitable, we propose a novel MDP formulation accompanied by a specific resolution algorithm.
\new{The fact of using MDP arises naturally since the Markov kernel corresponds to probabilistically quantified uncertainty on the contacts while the decisions (or the non-determinism) of the MDP correspond to the possibilities of routing decisions of each node at a given time.}
The \RU model is the first of its kind to consider \textit{rerouting}, which models both the fault detection and reaction time of the DTN routing agent.
Modeling this crucial and practical aspect allows us to introduce \IRU (a variation that uses only local information available on each node) and \CGRU (an extension to CGR that materializes routing under uncertain contact plans in existing DTN protocol stacks).
We evaluate and compare the \RU, \IRU and \CGRU in an appealing benchmark comprising networks with random failures as well as realistic case studies of Low-Earth Orbit (LEO) satellite networks with uncertain inter-satellite and ground contacts.
Results provide compelling evidence that \RU provides the adequate framework to route in uncertain DTNs.

\new{To summarize, contributions in this paper are enumerated as follows:
\begin{enumerate}
    \item We present a new uncertain DTN classification and model;
    \item We introduce \RU to route on uncertain DTNs based on a theoretical MDP formulation;
    \item We propose \IRU and \CGRU as concrete practical application approaches derived from \RU; and
    \item We evaluate \RU, \IRU and \CGRU in realistic fault-prone LEO satellite networks.
\end{enumerate}
}

The remaining of this paper is organized as follows.
Section~\ref{sec:model} presents the uncertain DTN network model which is used to construct the \RU model and derived \IRU and \CGRU in Section~\ref{sec:rucop}.
A comparison benchmark and subsequent results are presented, analyzed and discussed in Section~\ref{sec:analysis}.
Finally, the paper is concluded in Section~\ref{sec:conclusion}.

\section{Uncertain DTN Model}
\label{sec:model}

\subsection{Uncertain Time-Varying Graph}

In order to model a time-evolving and uncertain DTN network, the time-varying graph proposed in~\cite{Liang2017} is extended by uncertainty functions into an Uncertain Time-Varying Graph defined as follows. 

\textbf{Definition.} An Uncertain Time Varying Graph $\mathcal{G} = (G, \mathcal{T}, p_f, \varsigma, \fdd)$ is a Graph composed of the following components: 
\begin{enumerate}

    \item \textbf{Underlying (static) digraph $G = (V, E)$}.
    Represents the connectivity of the network that remains stable during a time slot. 
    
    \item \textbf{Time slot $\mathcal{T} \subseteq \mathbf{T}$}, where $\mathbf{T}$ is the time domain (e.g. the natural numbers).
    $\mathcal{T} = \{t_0, t_1,...,t_T\}$ is a discrete and finite time span set, where $T$ is an integer indicating the horizon of interest, measured in the number of slots.
    The slot length in $\mathcal{G}$ can be adjusted in order to capture (\textit{i}) the topological changes, and (\textit{ii}) the minimum period of time it takes a node to realize a link has failed to establish.

    \item \textbf{Edge failure probability function $p_f: E \times \mathcal{T} \rightarrow [0,1]$}.
    It indicates the probability an edge will not occur as expressed in the uncertain contact plan, i.e., a topology change respects the original schedule.
    Indeed, $p(e,t)=1-p_f(e,t)$, where $p(e,t)$ stands for the edge $e$ success probability at the time slot $t$. 
    A success probability of $p(e,t)=0$ indicates no contact is present at this edge.
    
    \item \textbf{Edge delay function $\varsigma: E \times \mathcal{T} \rightarrow \mathcal{T}$}. 
    It models the time data spend on crossing an edge between two nodes.
    When $\varsigma(e,t)=0$, the time is insignificant compared with the time slot duration, i.e., the data is delivered immediately.
    The value of the edge delay function stands for the number of time slots (i.e., $\varsigma(e,t)$ is an integer) required  for the target node to receive the traffic.
    
    \item \textbf{Edge failure detection delay function 
    $\fdd: E \times \mathcal{T} \rightarrow \mathcal{T}$}.
    It stands for the time it takes to detect a contact did not occur as expected.
    As with the edge delay function, $\fdd(e,t)$ is expressed as a number of time slots.
    In DTN protocol terminology, $\fdd(e,t)$ would represent the bundle custody acknowledge timeout. In general, $\fdd(e,t) \geq \varsigma(e,t)$.
\end{enumerate}

\begin{figure}
    \centering
    \includegraphics[width=0.6\linewidth]{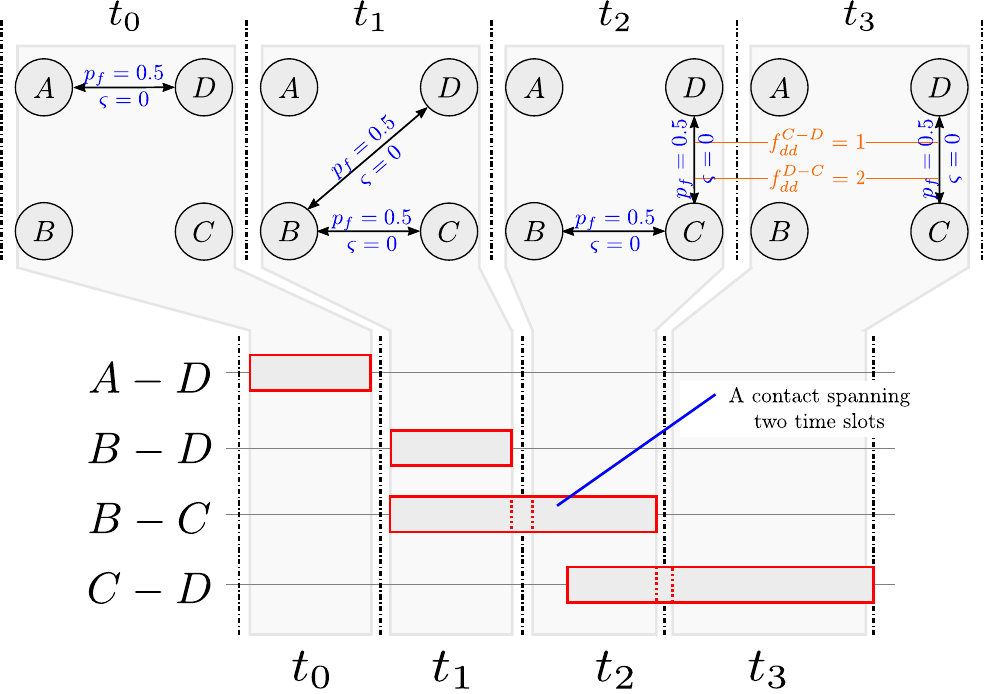}
    \caption{Uncertain time-varying graph model example with 4 nodes, 4 time slots $\mathcal{T}$ and 4 contacts.}
    \label{fig:model}
\end{figure}

Fig.~\ref{fig:model} illustrates an example DTN graph modeled by an uncertain time-varying graph.
All edges present in $G=(V,E)$ are configured with a failure probability function $p_f=0.5$ and a delay function $\varsigma=0$.
In the model, a contact between two nodes can span several time slots, such as the $B-C$ case spanning $t_1$ and $t_2$.
Also, a time slot can represent long and stable topological periods with the same underlying digraph, such as $t_3$ with an edge between $C-D$.
At $t_2$, node $C$ will be able to detect a failure on edge $C-D$ and react at the beginning of $t_3$, as its failure detection delay $\fdd^{C-D}=1$.
However, node $D$ will not do so before $t_3$ terminates since $\fdd^{D-C}=2$. 
Indeed, contacts in DTN are unidirectional and can have different properties on the forward and return link.

Failure probability $p_f$ in $\mathcal{G}$, $\varsigma$, and $\fdd$ are expressed on a per-slot basis.
Two modeling approaches with different interpretations are envisioned on this regard: coarse and fine grained slotting.

\minisection{Coarse-grained slotting:}
When time-slots are designed to contain full contacts (i.e., $B-D$ contact in $t_1$ in Fig.~\ref{fig:model}), then $p_f$ represents the failure probability of the whole contact. 
In other words, the whole contact exists or the whole contact fails.
In such case, an $\fdd=0$ would model the case where the failure of the contact is detected and reacted upon immediately at contact start time, while an $\fdd=1$ would represent the case where the contact is declared as failed only once it is finalized.
This approach is appropriate to model transient failures in nodes, for instance.
Also, coarse-grained slotting is particularly appealing for networks with sparse contacts, which can be bounded by a single time slot $t_n$ in $\mathcal{T}$.

\minisection{Fine-grained slotting:}
When a contact spans several smaller time slots (i.e., $B-C$ contact in $t_1$ and $t_2$ in Fig.~\ref{fig:model}), $p_f$ is the probability of failure of each of the slotted episodes comprising the contact.
In this case, a finer-grain slotting can be exploited to model independent transmission attempts within the contact.
An $\fdd=1$ would thus model a timeout equal to the bundle transmission duration and the round trip time delay for receiving a delivery confirmation.
Fine-grained slotting can be used to model contacts where poor channel conditions or interference from other sources render a successful transmission uncertain.

\subsection{Fault Detection and Rerouting}

Rerouting after effective detection of a failed contact or transmission attempt is a fundamental practical aspect to model the overall data flow in DTNs under uncertain contact plans.
Single route reliability estimations such as those in~\cite{Liang2017} can result inaccurate in practice when nodes detect and act upon unexpected failures.
However, the phenomena is not trivial.

\begin{figure}
    \centering
    \includegraphics[width=0.6\linewidth]{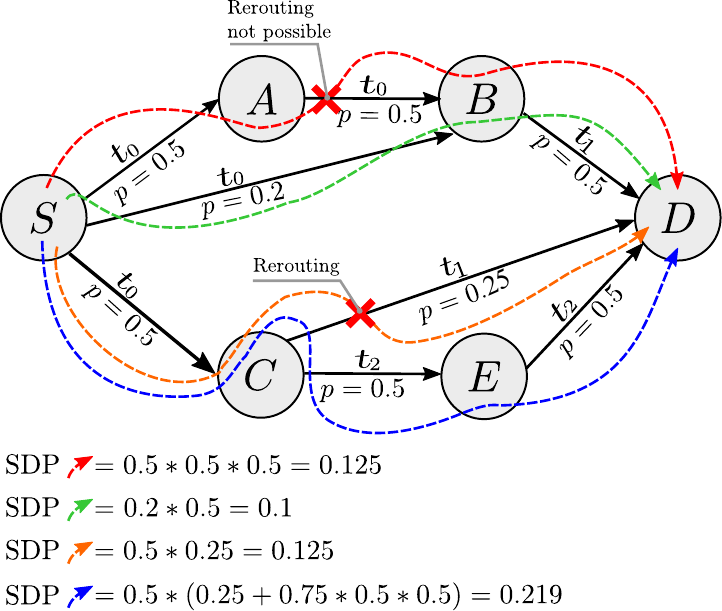}
    \caption{Rerouting is possible when node $C$ detects a failure at the end of $t_1$ ($\fdd=1$) and has an alternative route to $D$ at $t_2$ that arrives on the same time slot ($\varsigma=0$).}
    \label{fig:reroute}
\end{figure}

Consider the example of Fig.~\ref{fig:reroute} in which all links have a failure probability $p_f=0.5$ with the exception of $S \rightarrow B$  at $t_0$ and $C \rightarrow D$ at $t_1$ which have a failure probability of $p_f=0.80$ and $p_f=0.75$ respectively. 
The transmission delay $\varsigma=0$ and failure detection delay is $\fdd=1$ for all links and data flows from source $S$ to destination $D$.
Without considering rerouting, routes via node $A$ $(S \rightarrow A \rightarrow B \rightarrow D)$ and via node $C$ $(S \rightarrow C \rightarrow D)$ would be equally reliable because they both account for a \textit{successful delivery probability} (SDP) of $0.125$.
However, rerouting after failure detection might challenge this calculation.
If the link between $A \rightarrow B$ fails in the route via $A$, then the data will not reach the destination.
But, if the contact between $C \rightarrow D$ fails, it is still possible to relay the data to node $E$ after $t_1$, which has another route towards $D$.
In a context where rerouting is possible with $\fdd<=1$, the probability of a bundle to reach the destination via node C is 75\% higher ($\SDP=0.219$).
Otherwise, for $\fdd>=2$, the delivery probability through $C$ remains $SDP=0.125$.

In the following section, we claim the rerouting effect in an uncertain time varying graph can be properly represented by means of Markov Decision Processes.

\section{Routing Under Uncertain Contact Plans}
\label{sec:rucop}

\subsection{Markov Decision Process}

A Markov Decision Process (MDP) is a mathematical structure that
allows for the modelling of discrete-time systems in which the
interaction between non-deterministic and probabilistic behaviour is
central~\cite{Puterman:1994,FilarKoos:1996}.  
Thus, MDPs provide an
appropriate framework for modelling decision making on systems under
probabilistically quantified uncertainty.
%
In its simplest form, a MDP $\mdp$ is a tuple 
$(\St, \Act, \PM, s_0)$ where
\begin{itemize}
  \item $\St$ is a finite set of states with initial state $s_0\in\St$,
  \item $\Act$ is a finite set of actions, and
  \item $\PM: \St \times \Act \times \St \to [0, 1]$ is a transition probability function such that $\sum_{s' \in S} \PM(s, \alpha, s') \in \{0,1\}$, for all $s \in\St$ and $\alpha \in \Act$.
\end{itemize}
If $\sum_{s' \in S} \PM(s, \alpha, s')=1$, $\alpha$ is said to
be \emph{enabled} in $s$.  In this case, $\PM(s,\alpha,\cdot)$
can be interpreted as the probability distribution of choosing the next
state, conditioned to the fact that the system is in state $s$ and
action $\alpha$ has been chosen.  We notice that it is usually
required that at least one action is enabled in every state.
\new{Since the problem ahead is a reachability problem (instead of a cost or reward problem), the usual reward function does not play any role and hence we have omitted it in the definition of MDPs.}

The intuitive operational behaviour of the MDP $\mdp$ is as follows.
The computation of $\mdp$ starts at the initial state $s_0$.  Assume
now the computation has taken $n$ steps and reached state $s_n$.  At
this moment one of the enabled actions in $s_n$, say $\alpha_{n+1}$,
is chosen to resolve the non-determinism at this state.  The next
state $s_{n+1}$ is now sampled randomly according to distribution
$\PM(s_n,\alpha_{n+1},\cdot)$.

Different types of properties could be required to a MDP. The usual
objective is to find a \emph{policy} that maximizes or minimizes the
likelihood of the given property.  A \emph{policy} is a function
$\policy:\St\to\Act$ that defines the decision to be made in a
possible resolution of the non-determinism%
\footnote{Polices could be more complex, depending on the whole
  history rather than the current state, and selecting randomly among
  the enabled actions.  The definition given here correspond to the so
  called \emph{memoryless} and \emph{deterministic} policies, which is
  sufficient for our purposes.}.
Thus, limiting the MDP $\mdp$ to the choices of the policy $\pi$
defines a Markov chain for which probabilities can be calculated.

We are particularly interested on maximizing the probability to reach
a state in the set of \emph{goal states} $\GS\subseteq\St$ from the initial state $s_0$, say
$\Prmax_{s_0}(\reach(\GS))$. (In our case, $\GS$ is the set of states in which
bundles have been successfully delivered).  
Moreover, we want to
obtain the maximizing policy.
This problem can be solved using the Bellman equations as
follows~\cite{BaierK08}.  Let $\St^{=0}\subseteq\St$ be the set of
states whose probability of reaching a state in $\GS$ is 0. ($\St^{=0}$ could be
calculated in $\mathcal{O}(|\St|)$.)
For each state $s\in\St$, define
a variable $x_s$ which represents the maximum probability of reaching
a goal state in $\GS$ from $s$, that is $x_s=\Prmax_s(\reach(\GS))$.  Then,
precisely the vector $(x_s)_{s\in\St}$ is the unique solution of the
following equation system:
\begin{align*}
  x_s &= 1 && \text{if \ } s\in \GS \\
  x_s &= 0 && \text{if \ } s\in {\St^{=0}}\\
  x_s &= \max_{\alpha\in\Act(s)} \sum_{t\in S}\PM(s,\alpha,t)\cdot x_t
      && \text{if \ } s\in\St\backslash({\St^{=0}}\cup \GS)
\end{align*}
Besides, the maximizing policy $\pi^{\max}$ can be obtained as follows:
\begin{align*}
  \pi^{\max}(s) &= \argmax_{\alpha\in\Act(s)} \sum_{t\in S}\PM(s,\alpha,t)\cdot x_t
               && \text{if \ } s\in\St\backslash({\St^{=0}}\cup \GS)
\end{align*}
If $s\in{\St^{=0}}\cup \GS$, $\pi^{\max}(s)$ is not interesting as $s$ is
already a goal state, or it cannot reach it.

Reachability properties are standard properties in probabilistic model
checkers such as PRISM~\cite{Kwiatkowska2011}.  Indeed, we have
successfully modeled single-copy routing in DTNs under uncertain
contact plans in PRISM~\cite{Raverta2018} and derived optimal routes in
this case.
Unfortunately, PRISM cannot deal with the size of models we required,
specially when we consider DTNs with multiple copies.

\subsection{\RU}
\label{sec:rucop:rucop}

In order to determine the upper delivery probability bound for routing with $N$ copies in a DTN, we have developed Routing under Uncertain Contact Plans (\RU). 
\RU is an MDP formulation which encodes all possible routing decisions for an uncertain DTN network based on its uncertain time-varying graph representation and traffic parameters, comprising source, target and number of copies allowed.
This information is encoded in states and transitions.
Table~\ref{tab:notation} summarizes the notation used throughout the remaining of this section.


\begin{center}
\begin{longtable}{|p{4cm}|p{8.815cm}|}
\caption{Notation reference}
\label{tab:notation}\\
\hline
\multicolumn{1}{|c|}{\textbf{Symbol}} & \multicolumn{1}{c|}{\textbf{Description}} \\
\hline
\endfirsthead

\multicolumn{2}{|c|}{\textbf{Uncertain DTN Model (Section 2)}} \\ \hline
\arrayrulecolor{lightgray}
$p_f(e, t)$ & Failure probability for link $e$ at time slot $t$ \\ \hline
$\varsigma(e,t)$ & Delay for link $e$ at time slot t  \\ \hline
$\fdd(e,t)$ & Failure detection delay for link $e$ at time slot $t$ \\\hline
\arrayrulecolor{black}
$\mathcal{T}$ & Set of time slots \\ \hline

\multicolumn{2}{|c|}{\textbf{\RU Core Algorithm (Section~\ref{sec:rucop:rucop})}} \\ \hline
\arrayrulecolor{lightgray}
$G_{t_{i}}$ & Underlying digraph $G$ for time slot $t_{i}$\ \\ \hline
$\setSt_{t_{end}}$ & Set of successful final states  \\ \hline
$\setSt_{t_{i}}$ & Set of states at time slot $t_i$ \\ \hline
$\cp(c)$ & Number of copies at node $c$\\ \hline
$\mathcal{C}_{t_i}$ & Set of nodes carrying copies in time slot $t_i$ \\ \hline
$\pred^+_{G_{t_{i}}\!\!\!\!}(c)$ & Set of all nodes in $G_{t_{i}}$ reaching $c$ in at least one hop \\ \hline 
$\pth_{G_{t_i}}(c',c)$ & Set of directed path from $c'$ to $c$ in $G_{t_{i}}$ \\ \hline
$\setPaths_c$ & Set of paths leading to $c$ \\ \hline
$R$ & Set of rules (i.e. pairs of nr.\ of copies and a path) \\ \hline
$\mathcal{R}_c$ & Set of $c$-compatible sets of rules (i.e. set of rules transmitting exactly $\cp(c)$ copies from $c$) \\ \hline 
$\Tr(s)$ & Set of actions leading to state $s$ (an action is a set of rules distributing exactly $\mathit{num\_copies}$) \\ \hline
$\mathit{pr}_R$ & Successful probability of action $R$\\ \hline
$\Prob(s) $ & Successful delivery probability of state $s$ \\ \hline
$\SDP(R,s,t)$ & Successful probability for action $R$ starting from state $s$ at time slot $t$ (Algorithm \ref{Alg:sdp}) \\ \hline
$\textit{get\_prev\_state}(s, R)$ & Returns the state from which action $R$ leads to $s$\\ \hline
$\mathit{best\_action}(s)$ & The action from $s$ maximizing the delivery prob. \\ \hline
\arrayrulecolor{black}
$\RU(G, c, T)$ & Algorithm \ref{Alg:ru} \\
\hline

\multicolumn{2}{|c|}{\textbf{\RU SDP Computation (Section~\ref{sec:rucop:rucop})}} \\ \hline
\arrayrulecolor{lightgray}
$\powerset(X)$ & Power set of $X$ \\ \hline
$\mathit{contacts}(R)$ & Set of links involved in action $R$ \\ \hline
$\mathit{state\_af\_fl}(R, s, \mathit{fs})$ & Leading state when set of failures $\mathit{fs}$ happen \\ \hline
$\mathit{pr}_{\mathit{fs}}$ & Probability of all links in $\mathit{fs}$ failing\\ \hline
$\mathit{pr}_R $ & Successful delivery probability of action $R$ \\ \hline
\arrayrulecolor{black}
$\SDP(s)$ & Successful delivery probability of state $s$ \\ 
\hline

\multicolumn{2}{|c|}{\textbf{\IRU (Section~\ref{sec:iru})}} \\ \hline
\arrayrulecolor{lightgray}
$\mathit{Safe\_state}(n,c,\ts)$ &  State in which node $n$ has all $c$ copies available\\ \hline
$\LTr_{n}(\_,\_,\_)$ &  Routing table for node $n$ \\  \hline
\arrayrulecolor{black}
$\mathit{Post}(\LTr_{n}(\ts,\rc,\ts'))$ & The state known by node $n$ after action $\LTr_{n}(\ts,\rc,\ts')$ \\ 
\hline

\multicolumn{2}{|c|}{\textbf{\CGRU (Section~\ref{sec:rucgr})}} \\ \hline
\arrayrulecolor{lightgray}
$\Rl_n(\ts)$ & Set of partial routes computed by CGR at node $n$ for time slot $\ts$ \\ \hline
$\route$ & A partial route computed by CGR \\ \hline
$\route[i]$ & $i$th contact in the partial route $\route$ \\ \hline
$\Prob_n(\ts)$ & Prob. of delivering a copy from $n$ at time slot $\ts$ \\ \hline
$\mathit{src}(e)$ & Source of link $e$ \\ \hline
$\mathit{tgt}(e)$ & Destination of link $e$ \\ \hline
\arrayrulecolor{black}
$\SDPCGR(\route,\ts) $ & Bundle's delivery prob. through partial route $\route$ \\ \hline

\end{longtable}
\end{center}

\minisection{States}. Each state in \RU contains information of the number of copies present on each node in the network at a given time slot.
For example, in the network of Fig.~\ref{fig:reroute}, the initial state would be $s_{t_0}=[S^n A^0 B^0 C^0 D^0 E^0|t_0]$ denoting that $s_{t_0}$ has $n$ copies of the bundle at time $0$, the start time of $t_0$.
A state $s_{t_3}=[S^s A^a B^b C^c D^d E^e|t_3]$ at the beginning of $t_3$ would represent a successful delivery of data to $D$ as long as $d>=1$, meaning at least one copy of the data arrived at $D$ at the end of the time horizon.
Since it is assumed copies cannot be created or deleted, $s+a+b+c+d+e=n$ in all states.

\minisection{Transitions}. Transitions between states in \RU are composed by actions, which can be of two types: (\textit{i}) \textit{transmission transitions} imply a node perform a non-deterministic transmission through one (single-hop) or more edges (multi-hop) in $G$, and (\textit{ii}) \textit{store transitions} model the case where a node decides to keep the bundle in memory during the time slot.
Since state transitions imply a routing action on the nodes, the terms transitions and actions are used interchangeably in \RU.

\minisection{Tree Construction}.
To  build  the  state  and  transition  tree,  RUCoP  starts from the desirable \textit{successful states} where data was delivered to  the  destination.
Next,  it  considers  states  from  the previous time  slot  that  can  lead  to  the  current  state,  whether  by transmitting  data  through  a path  or  by  keeping  it  in storage. 
In order to determine which state of the previous time slot can arrive to the current state, a set of transmissions transition are constructed.
Finally, between these transitions, the one which has the highest delivery probability is chosen and noted.
The process repeats until the \textit{initial state} is reached.
In order to determine the probability of a given transition, all cases of failures and successful link establishments are considered: \begin{enuminline}
\item when a contact fails, data remains stored in the transmitting node where new transmission transitions can be considered after $\fdd$, and
\item when a link is established, the data is transmitted through it, and it can be sent again after $\varsigma$. 
\end{enuminline}

\begin{figure}
    \centering
    \includegraphics[width=0.7\linewidth]{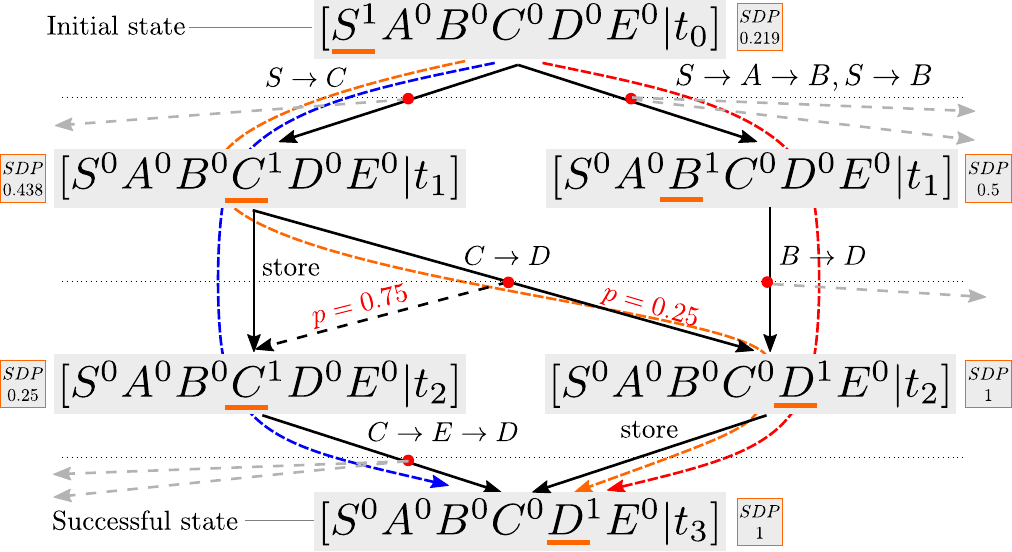}
    \caption{\RU MDP tree based on the network of Fig.~\ref{fig:reroute} for 1 copy.}
    \label{fig:rucop1}
\end{figure}

For example, the \RU model in Fig.~\ref{fig:rucop1} corresponds to the network of~Fig.~\ref{fig:reroute}, when a single copy is sent.
The successful state $[S^0 A^0 B^0 C^0 D^1 E^0|t_3]$ is at the last time slot $t_3$, which can be reached either by receiving data through $C \rightarrow E \rightarrow D$ (multi-hop transmission) or by having data already stored at $D$ since $t_2$.
In turn, these intermediate states can only be reached if a $C \rightarrow D$ transition or a $B \rightarrow D$ transition takes place on $t_1$.
It can be observed that, if $C \rightarrow D$ fails, $C$ can detect the failure ($\fdd=1$) and store the data for further transmission transitions.
However, if the contact $B \rightarrow D$ fails, data will remain in $B$ leading to state $[S^0 A^0 B^1 C^0 D^0 E^0|t_2]$, from which the successful state cannot be reached (i.e., delivery cannot occur).  
This is represented by the grey dotted arrow outgoing the red dot.
A similar (but more involved) situation happens in transitions $S \rightarrow A \rightarrow B$ and $S \rightarrow B$ outgoing the initial state: if the $S \rightarrow A$, $A \rightarrow B$ or $S \rightarrow B$ contacts fail, data will remain in $S$ leading to state $[S^1 A^0 B^0 C^0 D^0 E^0|t_1]$ or in $A$ leading to state $[S^0 A^1 B^0 C^0 D^0 E^0|t_1]$.
Both of these states are failure consequences of transitions $S \rightarrow A \rightarrow B$ and $S \rightarrow B$, which have no possibility of reaching the successful state (greyed-out arrows in the figure). 
In this simple example, all non-deterministic transmission transitions (red dots in the figure), except $C \rightarrow D$, lead to states unable to reach the successful state as long as some contact in the transition fails.
Indeed, constructing the tree backwards avoids exploring such states.
It is interesting to note that if detection delay would have been $\fdd=2$ in $C \rightarrow D$ at $t_1$, the dashed line indicating failure path would lead to $[S^0 A^0 B^0 C^1 D^0 E^0|t_3]$, which is also unable to reach the successful state.
In other words, by the time when $C$ detects the failure, the contact $C \rightarrow E$ would have already passed.

\minisection{Successful delivery probability.}
While constructing the tree, \RU keeps track of the successful delivery probability. 
Indeed, $\SDP=1$ at the successful states, and is updated as the tree is built backwards in time following the Bellman equations.
For each non-deterministic transmission transition, the probability of arriving to the successful state is computed.
$\SDP$ is updated with the highest probability.
Once the initial state is reached, the $\SDP$ will capture the maximum delivery probability possible. 
By navigating the tree top-down, the most reliable routing decisions (i.e., policy) can be obtained by choosing transitions that lead to states with the best $\SDP$ metric.
In the example, $S$ should route the data to $C$ at $t_0$ for an $\SDP=0.219$, and $C$ should try to send data to $D$ at $t_1$ for an $\SDP=1$, or to $E$ at $t_2$ in case of failure.

\begin{figure}
    \centering
    \includegraphics[width=0.7\linewidth]{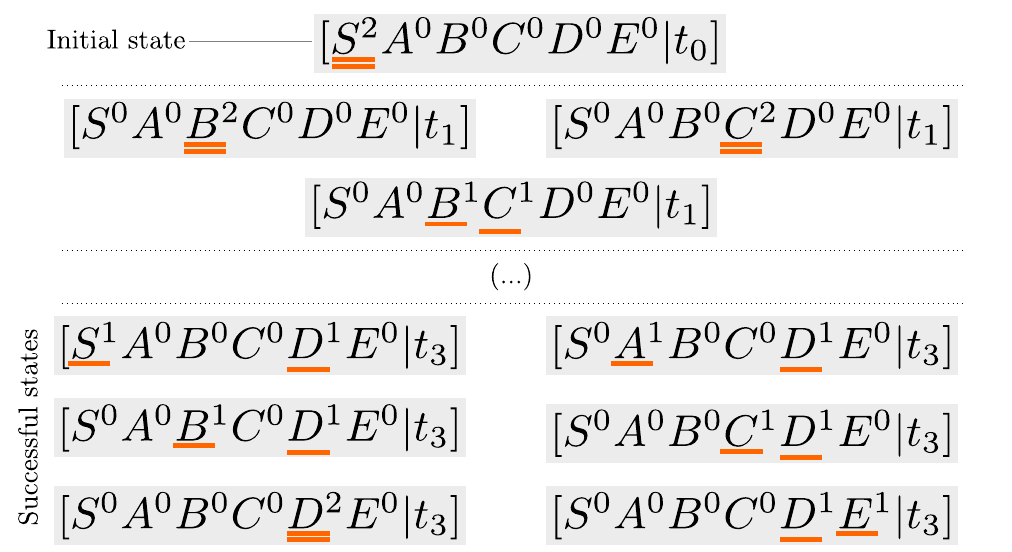}
    \caption{\RU MDP tree based on the network of Fig.~\ref{fig:reroute} for $2$ copies.}
    \label{fig:rucop2}
\end{figure}

\minisection{Multiple copies.} 
The proposed \RU expression is specifically designed to model the state of the network with multiple copies.
Naturally, modeling multiple copies notably increases the number of transitions and states in the MDP. For example, when two instances of the bundle are considered, transmission transitions can involve the transmission of either one or two bundles of data, and transmission failures might occur in any of the used links.
As illustrated in Fig.~\ref{fig:rucop2}, six successful states are possible and should be considered with two copies on the example network.
For instance, node $S$ can choose to transmit one copy via $A$ and one via $C$ to maximize the delivery chances.
However, for larger networks with several copies, constructing the model requires of the following formal expression of the \RU algorithm.

\minisection{The algorithm:}
For simplicity, we present the algorithm limited to uncertain time
varying graphs where the edge delay is insignificant and the edge
failure detection delay is always one time slot (i.e. $\varsigma(e,t)=0$
and $\fdd(e,t)=1$ for all edge $e\in E$ and time slot $t\in \mathcal{T}$). 
At the end of this section, we hint the required modifications of the algorithm to deal with the general treatment of these delays.
Algorithm \ref{Alg:ru} lists the formal steps required to construct
and solve the \RU MDP for these type of networks with a maximum of
$\mathit{num\_copies}$ number of copies.

\begin{algorithm}
  \begin{algorithmic}[1]
    \REQUIRE Uncertain time varying graph $\mathcal{G}$, $\mathit{num\_copies}$, Target
    \ENSURE Explored states $\setSt$, Routing table $\Tr$, Successful delivery probability $\Prob$
    \STATE determine \textit{successful states} $\setSt_{t_{end}}$ for $\mathit{num\_copies}$
    \STATE $\setSt \gets \setSt_{t_{end}}$
    \FORALL {$t_i \in  \mathcal{T}$, starting from $t_{end-1}$}
      \STATE $\setSt_{t_{i}} \gets \varnothing$
      \FORALL {state $s \in \setSt_{t_{i+1}}$}
        \STATE determine \textit{carrier nodes} $\setCN_{t_{i}}$
        \FORALL {node $c \in \setCN_{t_{i}}$}
          \STATE $\setPaths_c \gets \{c\}\cup\bigcup_{c'\in\pred^+_{G_{t_{i}}\!\!\!\!}(c)} \pth_{G_{t_{i}}\!\!\!\!}(c',c)$
          \STATE $\setRules_c \gets \big\{ {R\subseteq \{0,\ldots\cp(c)\}\times\setPaths_c} \mid \ {\sum_{(k,\rho)\in R} k = \cp(c)}\big\}$
        \ENDFOR
        \STATE $\Tr(s) \gets \big\{ \bigcup_{c\in\setCN_{t_{i}}} R_c \mid \forall {c\in\setCN_{t_{i}}}: R_c \in \setRules_c\big\}$%
        \FORALL{$R \in \Tr(s)$}
          \STATE $s' \gets \textit{get\_previous\_state}(s, R)$
          \STATE $\setSt_{t_{i}} \gets \setSt_{t_{i}}\cup\{s'\}$
          \STATE $\mathit{pr}_R \gets \SDP(R,s',t_{i})$
          \IF{$\Prob(s')$ is undefined or $\Prob(s') < \mathit{pr}_R$}
            \STATE $\Prob(s') \gets \mathit{pr}_R$
            \STATE $\textit{best\_action}(s') \gets R$
          \ENDIF
        \ENDFOR
        \STATE $\setSt \gets \setSt \cup \setSt_{t_{i}}$
      \ENDFOR
    \ENDFOR
    \RETURN $\setSt$, $\Tr$, $\Prob$
  \end{algorithmic}
  \caption{The \RU algorithm} \label{Alg:ru}
\end{algorithm}

Initially, a set of all possible \textit{successful states} $\setSt_{t_{\mathit{end}}}$ are generated (line 1) and added to the set of explored states (line 2).
A state is successful if at least one copy is in the target node and exactly $\mathit{num\_copies}$ are distributed among all nodes.
\RU builds the MDP backwards from this set with the goal of arriving to the initial state.
To this end, all reachable states $\setSt_{t_{i}}$ within each $t_i$ in $\mathcal{T}$ are determined starting from an empty set (line 4 and loop starting at line 5).
$\setSt_{t_{i}}$ is subsequently populated with all states that are able to reach some state in $\setSt_{t_{i+1}}$ by means of actions involving bundle transmissions, data storage or a combination of them when multiple copies are presented.

Thus, for each state $s \in \setSt_{t_{i+1}}$, the loop proceeds in two parts.  The first one (lines 6-11) determines the set of actions $\Tr(s)$ that successfully lead to state $s$.  The second one (lines 12-20) calculates the predecessor states for each of these actions which are then included in the set of states $\setSt_{t_{i}}$ of the preceding time slot and for which its successful delivery probability (SDP) is calculated.

To obtain $\Tr(s)$, the set of \textit{carrier nodes}
$\setCN_{t_{i}}$ in $s$ is first determined (line 6).  A
carrier node is a node holding at least one copy of the bundle.
An action in $\Tr(s)$ is a set of \emph{rules}. A rule is a tuple $(k, \rho)$ where
  $\rho$ is a valid single-hop or multiple-hop path (or route) in the underlying
  digraph $G$ for the time slot $t_{i}$ ($G_{t_{i}}$), and 
  $k$ is the number of copies transmitted through this path; thus,
  $k\leq \cp(c)$, where $\cp(c)$ is the number of copies the target
  carrier node $c$ has in its buffer.

For each carrier node $c\in \setCN_{t_{i}}$, the set
$\setPaths_c$ of paths leading to $c$ in the current contact digraph
$G_{t_{i}}$ is determined.  This is calculated in line~8 where:
\begin{enumerate*}[label=(\roman*)]
\item%
  $\pred^+_{G_{t_{i}}\!\!\!\!}(c)$ is the set of all nodes
  in $G_{t_{i}}$ reaching $c$ in at least one hop, and
\item%
  $\pth_{G_{t_{i}}\!\!\!\!}(c',c)$ is the set of all paths in
  $G_{t_{i}}$ starting in node $c'$ and ending in $c$ containing all
  distinct vertices.
\end{enumerate*}
In addition, $\setPaths_c$ always contains the trivial path
$c$ which is intended to represent that data remains stored
in the node $c$ for the current time slot.

Notice that the different copies may arrive at node $c$ through
multiple paths.  Thus $\setRules_c$ contains the set of all
\emph{compatible} sets of rules that indicate how the copies arrive to
$c$ (line~9).  By compatible, we mean that the numbers of copies
delivered by the rules in such set should add up to exactly $\cp(c)$,
i.e., $R\in\setRules_c$ whenever $\sum_{(k,\rho)\in R} k = \cp(c)$.

Finally (line~11), an action $R\in\Tr(s)$ is a set of rules so that,
for each carrying node $c\in\setCN_{t_{i}}$, the subset of
all rules in $R$ leading to $c$ is compatible (i.e.,
$R\cap(\mathbb{N}\times\setPaths_c)\in\setRules_c$).
A rule $R$ never delivers more than $\mathit{num\_copies}$ in
total. This is guaranteed by the fact that
$\sum_{c\in\setCN_{t_{i}}}\cp(c) \leq \mathit{num\_copies}$.

\begin{figure}
    \centering
    \includegraphics[width=\linewidth]{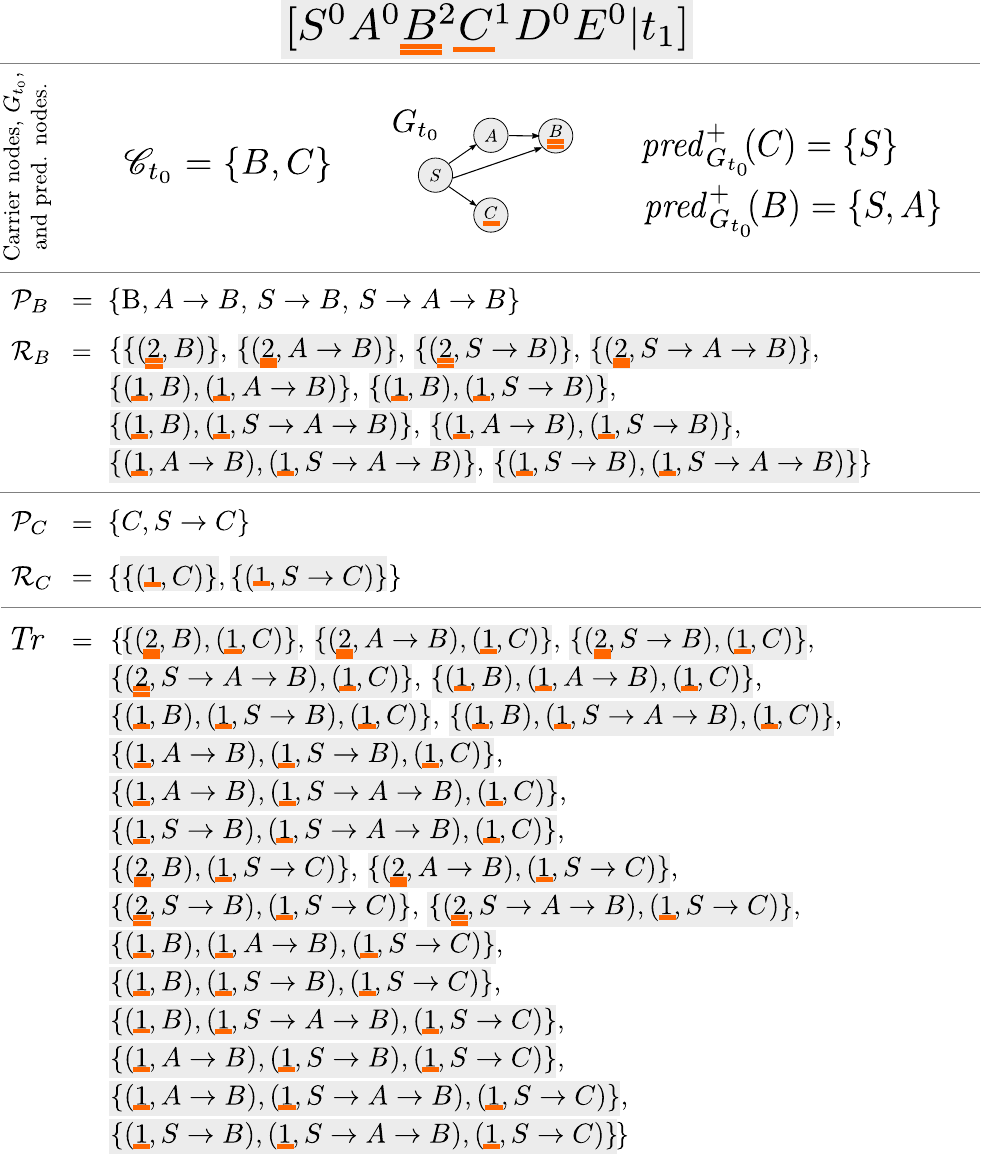}
    \caption{Nodes, rules and transition example in \RU}
    \label{fig:rules}
\end{figure}

To illustrate the exposed concepts, Fig.~\ref{fig:rules} lists carrier 
and predecessor nodes, paths, rules and transition for state
$s = [S^0 A^0 B^2 C^1 E^0 D^0 | t_1]$ corresponding to the network in
Fig.~\ref{fig:reroute} when 3 copies are allowed.  Since $B$ carries
two copies, each compatible set of rules leading to $B$ may have up to
two rules.  The resulting transition includes all the possible
transmissions of the copies succesfully reaching the evaluated state.

Each transition $R \in \Tr(s)$ is considered individually to determine
its corresponding previous state $s'$ (line 13) which is added to the
set of previous sates $\setSt_{t_{i}}$ (line 14).  Notice that $s'$
may already be present in $\setSt_{t_{i}}$ if it is the source of a
previously analysed transition $\hat{R} \in \Tr(\hat{s})$ for some
previously selected state $\hat{s}\in\setSt_{t_{i+1}}$.
In line 15, the probability induced by transition $R$ is calculated
calling function $\SDP$ (which we will shortly discuss).
%
If this is the first time state $s'$ is visited (hence its successful deliver probability $\Prob(s')$ is not yet defined) or its previously
assigned probability is smaller than the newly found $\mathit{pr}_R$
(line 16) $\Prob(s')$ is set to the new maximum $\mathit{pr}_R$ (line
17) and indicated that this is achieved through transition $R$ (line
18). (This is implementing the maximum of the Bellman equations.)
Finally, all states explored at time slot $t_{i}$ are added to the set of of explored states $\setSt$ (line 21).
The next iteration will explore the new set of states $\setSt_{t_{i}}$
and so forth until $t_0$ is reached.

If the initial state $s_{t_0}$ --where all copies are present at the
source node-- is part of the set of explored states $\setSt$, then there is a series
of actions (stored in array $\textit{best\_action}$) that lead to a
successful delivery of the data with an optimal SDP equal to
$\Prob(s_{t_0})$.  If the initial state $s_{t_0}$ is not present in
$\setSt$, then $\Prob(s_{t_0})$ is undefined and the SDP for the
model is $0$, implying no routing decision can be successful in
delivering the bundle of data to the intended destination.

\minisection{Calculating SDP.}
Algorithm~\ref{Alg:sdp} shows how SDP is computed for a transition
$R$ leaving a state $s$.
We let $\mathit{contacts}(R)$ be a set containing every link involved
in some path in $R$, and iterate for every possible combination of
link failures (line 2).  Thus, a failure set $\mathit{fs} \in
\powerset(\mathit{contacts}(R))$ stands for a set of links that failed
to be established whereas $\mathit{contacts}(R) - \mathit{fs}$ are the
links that successfully transmitted the data.
Depending on $\mathit{fs}$, a transition comprising several hops can
leave the bundle in different nodes in the path and thus lead to
different states.  The state $\mathit{to\_state}$ to which the network
would evolve to if links in $\mathit{fs}$ failed is computed (line 3).
Notice that $\mathit{to\_state}$ may not be a successfully delivering
state in which case $\SDP(\mathit{to\_state})$ will not be defined and
the probability of delivering of this particular combination of
failing links is 0.  The conditional statement of line 4 takes this
into account. Thus, if $\mathit{to\_state}$ is a successful delivering
state, the probability $\mathit{pr}_{\mathit{fs}}$ of this failure set
to happen is calculated (line 5) and the contribution to the total
probability of successfully delivering the data when links in
$\mathit{fs}$ fail is added up (line 6).

\begin{algorithm}
  \begin{algorithmic}[1]
    \REQUIRE Transition $R$, state $s$, current time slot $t$
    \ENSURE SDP of current action
    \STATE $\mathit{pr}_R \gets 0$
    \FORALL{$\mathit{fs} \in \powerset(\mathit{contacts}(R))$}
      \STATE $\mathit{to\_state} \gets \mathit{state\_after\_failures}(R, s, \mathit{fs})$
      \IF{$\SDP(\mathit{to\_state})$ is defined}
        \STATE $\mathit{pr}_{\mathit{fs}} \gets \left(\prod_{e \in \mathit{contacts}(R) {-} \mathit{fs}} 1 {-} p_f(e,t)\right) * \left(\prod_{e \in \mathit{fs}} p_f(e,t)\right)$
        \STATE $\mathit{pr}_R \gets \mathit{pr}_R + \mathit{pr}_{\mathit{fs}} * \SDP(\mathit{to\_state})$
      \ENDIF
    \ENDFOR
    \RETURN $\mathit{pr}_R$
	\end{algorithmic}
	\caption{Successful Delivery Probability ($\SDP$)} \label{Alg:sdp}
\end{algorithm}

\begin{figure}
    \centering
    \includegraphics[width=\linewidth]{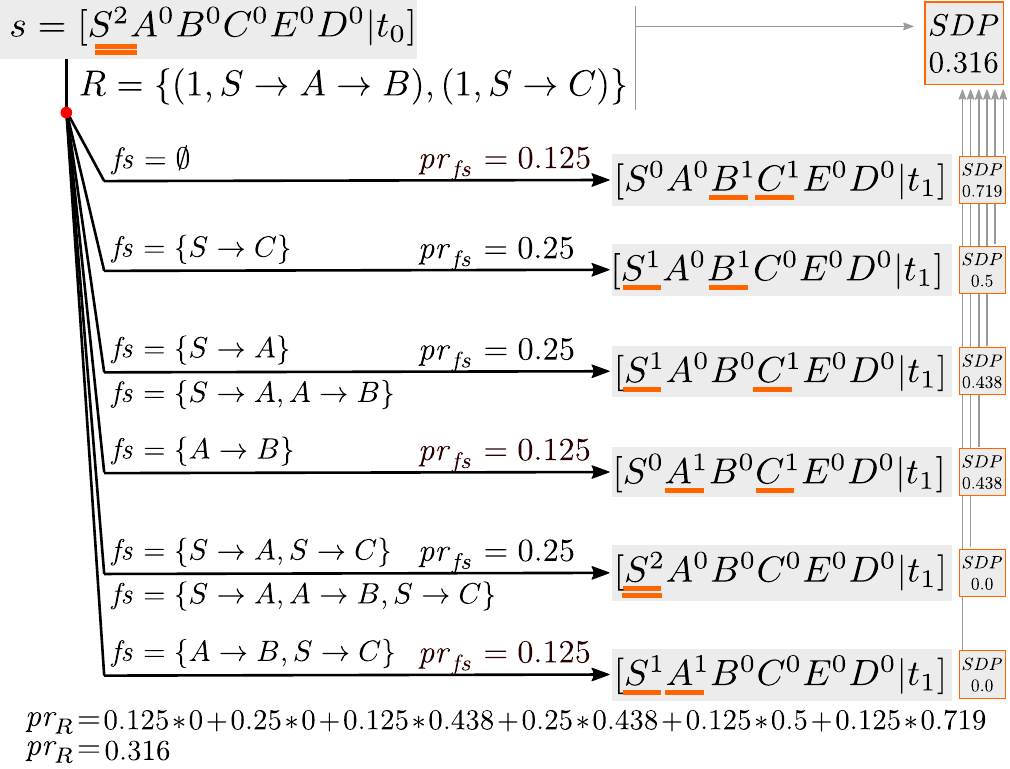}
    \caption{\RU updates for transition $\{(1, S \rightarrow A \rightarrow B), (1, S \rightarrow C)\}$ in-going state $[S^0 A^0 B^1 C^1 E^0 D^0 | t_1]$
    }
    \label{fig:sdp}
\end{figure}

Fig.~\ref{fig:sdp} illustrates the calculation of the SDP for transition $\{(1, S \rightarrow A \rightarrow B), (1, S \rightarrow C)\}$ which is a transition from $[S^2 A^0 B^0 C^0 E^0 D^0 | t_0]$ (the initial state) to  $[S^0 A^0 B^1 C^1 E^0 D^0 | t_1]$ when 2 copies are allowed and successfully transmitted.
In other words, when no failure is observed ($\powerset=\emptyset$), copies are successfully transmitted to $B$ and other to $C$ with a probability of $p=5^3=0.125$.
However, different failures can lead to $5$ possible alternative states with an accumulated probability of $1-0.125$.
Two of these have an undefined SDP, implying they have no further possibility of delivering the data to the destination.
This particular transition is the one with the highest SDP for $[S^2 A^0 B^0 C^0 E^0 D^0 | t_0]$ so that it stands for the optimal decision for forwarding two copies from $S$ to $D$ in the example network.

\minisection{Complexity analysis.}
First of all, notice that, if $N_c = {|\pred^+_{G_{t_{i}}\!\!\!\!}(c)|}$, then 
${|\setPaths_c|} \ \leq \ {N_c!\cdot\sum_{i=0}^{N_c}\frac{1}{i!}} \ < \ {e N_c!}$
and hence
${|\setRules_c|} \ \leq \ \binom{{|\setPaths_c|}+{\cp(c)}}{\cp(c)} \ < \ \binom{{e N_c!}+{\cp(c)}}{|\cp(c)}$.
From this, we have that
\[{|\Tr(s)|} \ = \ {\prod_{c\in\setCN_{t_{i}}} {|\setRules_c|}}
             \ < \ {\prod_{c\in\setCN_{t_{i}}} \binom{{e N_c!}+{\cp(c)}}{\cp(c)}}
             \ \leq \ {\binom{{e N!}+{K}}{K}}^K.\]
The last inequality follows from taking the worst case values, knowing
that $\cp(c)\leq\mathit{num\_copies}$ and
${|\setCN_{t_{i}}|}\leq\mathit{num\_copies}$ (there can never be more
carrier nodes than allowed copies), and letting
$N=\max_{t\in\mathcal{T}}\max_{c\in\setCN_{t}} N_c$ and
$K=\mathit{num\_copies}$.
The calculation of $\setPaths_c$ is done by a search algorithm of
complexity $\bigO{N_c!}$, and the construction of $\setRules_c$ and
$\Tr(s)$ are by enumeration. Thus, the complexity of lines 5-10 in Algorithm~\ref{Alg:ru} is $\bigO{{\binom{{e N!}+{K}}{K}}^K}$.

Focusing now in Algorithm~\ref{Alg:sdp}, notice that
$\mathit{contacts}(R)$ can contain, in the worst case, all edges
present in $G_{t_i}$;
therefore $|\mathit{contacts}(R)|\leq N_c^2 \leq N^2$.
Calculation in line 5 involves a multiplication of
$|\mathit{contacts}(R)|$ terms.  Hence, taking into account that the
loop repeats $|\powerset(\mathit{contacts}(R))|$ times, the complexity
of this algorithm is $\bigO{N^2 2^N}$.

From the previous observation, we see that the body of loop in lines
4-20 in Algorithm~\ref{Alg:ru} is
$\bigO{N^2 2^N {\binom{{e N!}+{K}}{K}}^K}$.
By observing that that $|\setSt_{t_i}| = \binom{{|V|}+K}{K}$, we can
finally conclude that the complexity of Algorithm~\ref{Alg:ru} is:
\begin{center}
$\bigO{N^2\cdot 2^N\cdot {\binom{{e N!}+{K}}{K}}^K\cdot \binom{{|V|}+K}{K}\cdot {|\mathcal{T}|}}$
\end{center}
Where $V$ is the set of all nodes in the network and
$\mathcal{T}$ is the time span under consideration.
We remark that, although in the worst case $N={|V|}$, we normally
expect $N$ ---the maximum number of nodes reaching a carrier node in a
single time slot--- to be significantly smaller than the number of
nodes in $V$.

Taking into account Stirling's aproximation to factorials, we finally
notice that the algorithm is in 2-EXPTIME.
However, in practice, we manage to have a satisfactory performance in practical use cases as it can be seen in Section~\ref{sec:discussion}.

\minisection{Link and failure detection delays.}
Algorithm~\ref{Alg:ru} is presented for networks with insignificant
link delays and one time slot failure detection delay in all cases.
In the general case, for networks where $\varsigma(e,t)>0$ or
$\fdd(e,t)>1$, for some link $e\in E$ and time slot $t\in\mathcal{T}$,
additional bookkeeping is necessary.  In particular, it is not
possible to only count copies of bundles. In this case, it will be
necessary to distinguish each copy and annotate it with the time slot
in which it is available for transmission (either because of the delay
after transmission, or because of the delay after failure).
This will have to be carefully considered, especially, when
calculating $\pth_{G_{t_{i}}}(c,c')$ (line 7 in
Algorithm~\ref{Alg:ru})
or the target state in
$\mathit{state\_after\_failures}(R,s,\mathit{fs})$ (line 3 in
Algorithm~\ref{Alg:sdp}).
In addition, this modification will have an impact on the (already
high) complexity of the algorithm.


\subsection{\IRU}
\label{sec:iru}

\begin{figure}
    \centering
    \includegraphics[width=0.8\linewidth]{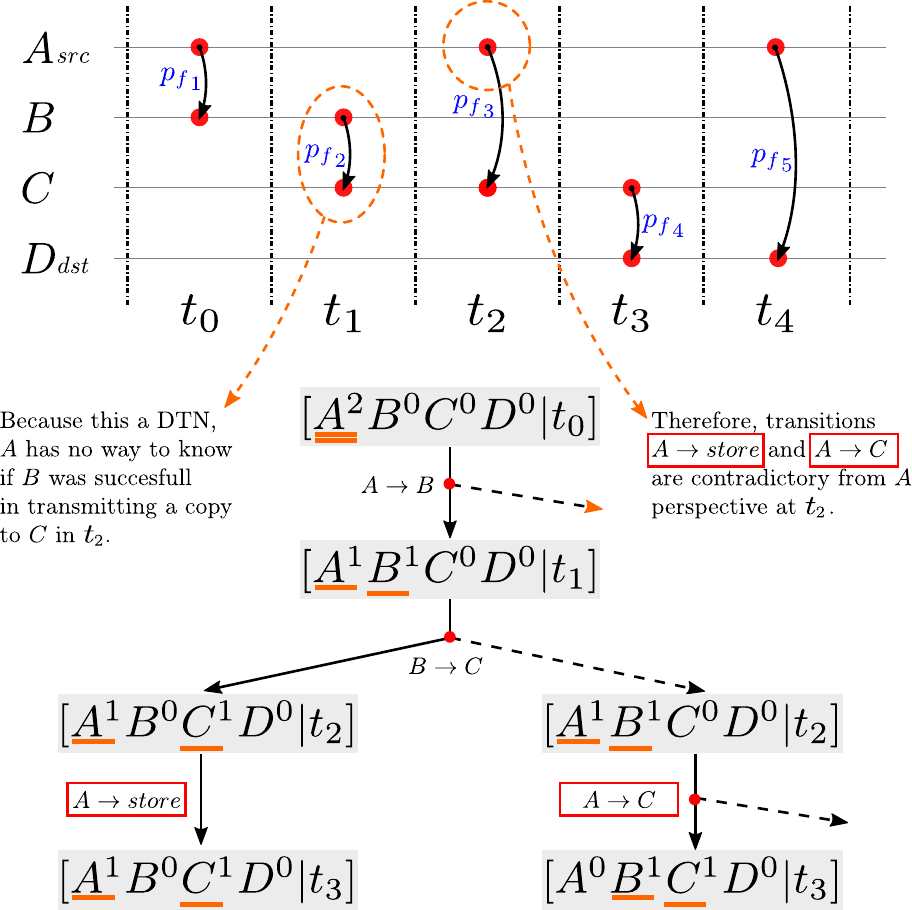}
    \caption{An example where local knowledge on $A$ is not enough to determine the global status of the system.}
    \label{fig:realizability}
\end{figure}

\RU is based on a global view of the system: decisions are taken based on the current state of the network.
This implies that distributed nodes need to know where all copies are in the network at any moment, including remote and potentially disconnected nodes.
Although optimal, this is impossible to achieve in highly partitioned DTNs where delays and disruptions force nodes to decide based on partial local knowledge~\cite{eddy1996hidden,tcs/CheungLSV06,tcs/GiroDF14}.
A simple example of this phenomenon is presented in Fig.~\ref{fig:realizability}.
Two decisions are possible at node $A$ in $t_2$, it can \textit{store} the copy or forward it to $C$.
However, which is optimal, might depend on weather the other copy is on $B$ or $C$ at $t_2$ (and also on ${p_f}_4$ and ${p_f}_5$).
Nonetheless, because $A$ was out of reach of $B$ and $C$, or because the contact $A-C$ is unidirectional or highly delayed, node $A$ may not be able to know which is the global status of the system nor which is the optimal action in $t_2$.
The aim of this section is to propose a derivation of \RU that can be implementable in DTNs where knowledge is restricted to each node's local view.
We coin this practical approach \textit{local} \RU (\IRU).

\IRU takes routing decisions on each local node $n$ using a pre-filled routing matrix $\LTr_{n}(t_s,c,t_i)$.  In this entry, $t_s$ indicates the ``safe'' time slot and it is normally the next one after the copies have been received, $c$ is the current number of copies that $n$ holds, and $t_i\geq t_s$ is the current time slot.
$\LTr_{n}(t_s,c,t_s)$ will contain the best decision $n$ can take assuming no knowledge of the network.  This is the same as if assuming that $n$ holds all copies and no other copy is in the system. Therefore $\LTr_{n}(t_s,c,t_s)$ contains exactly all routing decisions made by \RU for the state in which $n$ contains all $c$ copies and no copies are in the other nodes.
Nonetheless, if $n$ decides to keep some copies $\rc<c$ and only send $c-\rc$ copies, in the following time slots $n$ has certain knowledge of the previously distributed copies that may be handy to improve the decision on the routing of the remaining $\rc$ copies.
We illustrate this peculiarity using the contact plan in Fig.~\ref{fig:realizability} assuming that ${p_f}_1={p_f}_2=0.1$,  ${p_f}_3={p_f}_4=0.5$ and ${p_f}_5=0.9$.
The optimizing route for $\LTr_{A}(t_2,1,t_2)$, in which $A$ has no knowledge of the past, is to deliver the only copy through node $C$ with a probability of success of $0.25$ (the probability of success if delivering later directly to $D$ is $0.1$).
However, if $A$ had delivered a copy at time slot $t_0$ and preserved a second copy, the optimizing route for $\LTr_{A}(t_0,1,t_2)$ would be to keep the copy and deliver it later through $D$ (with probability $0.4645$, against $0.4525$ if the second copy is delivered through $C$ instead).


\IRU considers this peculiarity to optimize the decisions.  This means that populating the matrix requires $N$ different executions of \RU. 
Since nodes in DTN networks may not have powerful on-board computers, a centralized node, such as the mission operation and control (MOC) center in the case of satellite networks, should be responsible for computing $\LTr_{n}(t_s,c,t_i)$ and providing it to the network nodes in advance.


\begin{algorithm}
  \begin{algorithmic}[1]
    \REQUIRE number of copies $N$, target node $T$
    \ENSURE A routing table $\LTr_n$ for each node $n$
    \FORALL {$c \leq N$}
        \STATE $(S_c, \Tr_c, \Prob_c) \gets \RU(G, c, T)$
    \ENDFOR
    
    \FORALL {node $n$, time slot $\ts$, and $c \leq N$}
        \STATE $s \gets \mathit{Safe\_state}(n,c,\ts)$
        \IF {$s \in S_c$}
            \STATE {$\LTr_{n}(\ts,c,\ts) \gets \{ {(k,r) \in \Tr_c(s)} \mid  {\mathit{first}(r) = n} \}$}
            \STATE $\ts' \gets \ts$
            \STATE $\rc$ $\gets$ $(\exists$ $(k,n) \in LT_r(n,\ts,c,\ts')$$)?$ $k : 0$
            \WHILE{$\rc > 0$}
                \STATE $s' \gets \mathit{Post}(\LTr_{n}(\ts,\rc,\ts'))$
                \STATE $\ts' = \ts' + 1$
                \IF {$s' \in S_{\rc}$}
                    \STATE  $\LTr_{n}(\ts,\rc,\ts') \gets \{(k,r)  \in {\Tr_{\rc}(s')} \mid {\mathit{first}(r) = n} \}$
                \ELSE
                    \BREAK
                \ENDIF
                \STATE $\rc \gets (\exists$ $(k,n)  \in \LTr_{n}(\ts,\rc,\ts')) ?$ $k : 0$
            \ENDWHILE
        \ENDIF
    \ENDFOR
    \RETURN $\LTr_n$, for all node $n$.
	\end{algorithmic}
	\caption{\IRU Route table construction} \label{Alg:ltr}
\end{algorithm}

The construction of the \IRU matrix is detailed in Algorithm~\ref{Alg:ltr}.
First, \RU is executed for all possible $c \leq N$ copies, storing the resulting states, transitions and delivery probabilities $(S_c, \Tr_c, \Prob_c)$ (lines 1-2).

Notice that at this point all possible optimizing decisions have been calculated.  So, what remains of the algorithm, is to construct all tables $\LTr_{n}$ by properly searching on the results calculated with \RU.
Thus the algorithm nests two loops.  The outer loop (lines 4-21) iterates on  every node $n$, time slot $\ts$, and number of copies $c\leq N$ in order to first calculate the ``safe'' decision $\LTr_{n}(\ts,c,\ts)$. If needed, it then iterates on the inner loop (lines 10-19) to populate the table entries $\LTr_{n}(\ts,\rc,\ts')$ on the following time slots $\ts'>\ts$ for the distribution of the copies that have been held by the node.

So, the first step of the outer loop is to define the state $s$ in which the node $n$ has all copies $c$ in time slot $\ts$ (line 5) and no other copy is in the network. Thus $\mathit{Safe\_state}(n,c,\ts) = [A_0, B_0, ..., n_c, ... | \ts]$.  This is the ``safe'' state in which $n$ has no knowledge of the network.
If this state exists in $S_c$ (i.e. the corresponding \RU found a likely successful route to the target node), node $n$ has a route to target and its routing decisions (calculated through $\Tr_c(s)$) are saved in $\LTr_{n}(\ts,c,\ts)$ (line 7).
At this point, the number of copies $\rc$ that are not distributed in this routing action is calculated (line 9) and the current time slot $\ts'$ is set to $\ts$ (line 8).  If some copy remains in the node, the inner loop takes action (line 10).
Firstly, the state $s'$ known by node $n$ after taking the last routing decision (namely, $\LTr_{n}(\ts,\rc,\ts')$) is calculated (line 11).
More precisely $\mathit{Post}(\LTr_{n}(\ts,\rc,\ts'))$ delivers the state at time slot $\ts'+1$, in which node $n$ contains the copies remaining after routing action $\LTr_{n}(\ts,\rc,\ts')$, any node $n'$ that is in direct contact with $n$ --according to $\LTr_{n}(\ts,\rc,\ts')$-- contains exactly the number of copies that $n$ delivered to it, and any other node does not contain any copy.
Also, the next time slot is calculated (line 12).
If state $s'$ exists in $S_{\rc}$ (i.e. the corresponding \RU found a likely successful route to the target node), the routing decision is saved (lines 14).  Instead, if $s'$ was not marked as explored by \RU, then no path to the successful state is possible from $s'$, the action for that table entry is left undefined (line 16) and the inner loop is finished.
While there is a successful route to the target node, the number of remaining copies $\rc$ for the next step are calculated (line 18) and the inner loop repeats until no further copies $\rc$ remains in $n$.


It is worth to recall that $\LTr_{n}(\ts,c,\ts)$ is always the \textit{safe entry} to look up for the local node.
This means that whenever new copies arrive, or a routing decision fails to be accomplished in node $n$, it should take the current time slot $\ts$ as a safe place and look up the table at entry $\LTr_{n}(\ts,c,\ts)$ (assuming $c$ is the current number of copies held by $n$).
Because of this fact of returning to the ``safe entry'' each time of uncertainty, in which the node assumes no copies are present in remote nodes, \IRU accounts for a pessimistic-case knowledge from the local node perspective.
Nevertheless, we show in Section~\ref{sec:analysis} that \IRU is a valuable routing approach for uncertain contact plan implementable in realistic DTN nodes constrained to localized knowledge.

\subsection{\RU-enhanced CGR}
\label{sec:rucgr}

To easily exploit the \RU method in existing DTN protocol stacks with minimal modifications, we also propose an alternative CGR formulation (a single-copy DTN routing scheme). 
We base the approach on a \RU-based SDP metric to achieve reliably delivery of bundles over an uncertain contact plan.
CGR is a Dijkstra-based distributed routine that runs on each DTN node to determine the best routes to a given destination based on a pre-provisioned contact plan (the interested reader can refer to~\cite{Araniti2015},~\cite{FRAIRE2021102884} and~\cite{Fraire2017-Hindawi} for an in-depth description of CGR). 
We propose \CGRU as a simple means of extending CGR to operate with uncertain contact plans based on the outcomes of \RU.  The idea is that \CGRU selects the route that optimizes the successful delivery probability (SDP) instead of optimizing the time to destination as it is normally done in CGR.

In \CGRU, we let CGR calculate the list of possible routes to a given destination using its modified Dijkstra contact plan search.
In other words, route computation is left unchanged from legacy CGR.
Also, the resulting route list for each destination is constructed and consulted on forwarding time by the DTN node.
However, instead of choosing the best route from the list based on the best delivery time metric, \CGRU decides considering a custom SDP-based metric.
\CGRU metric is built around the $\Prob$ table constructed in Algorithm~\ref{Alg:ru} for only 1 copy.  
More precisely, for each node $n$ and time slot $\ts$, we take $\Prob_n(\ts)=\Prob(\mathit{Safe\_state}(n,1,\ts))$  ($\mathit{Safe\_state}$ is defined as in Sec.~\ref{sec:iru}). 
That is $\Prob_n(\ts)$ is the probability of successfully delivering a single copy from node $n$ at time $\ts$.
Similarly to \IRU, the values of $\Prob_n(t_s)$ can be pre-computed and provisioned to the DTN nodes together with the contact plan required by CGR to operate.

For the calculations, we assume that, after running CGR, a node $n$ is left with a table $\Rl_n:\mathcal{T}\to\powerset(E^*)$ that, given a time slot $\ts$, returns a set of partial routes $\Rl_n(\ts)$.  Each $\route\in\Rl_n(\ts)$ is a sequence of contacts --recall that each contact is an edge $e\in E$ of the uncertain timed-varying graph-- representing a partial route to destination, more precisely, the fragment of the route that starts in node $n$ at time slot $\ts$ and contains all hops that take place only during the same time slot.
Thus, for instance, considering the graph of Fig.~\ref{fig:reroute}, $\route = (S\to A)\,(A\to B)$ is a possible route in $\Rl_S(t_0)$, but $(S\to A)\,(A\to B)\,(B\to D)$ is not, as it expands through two time slots ($t_0$ and $t_1$), nor is $(S\to A)$, since it does not contains all the hops in time slot $t_0$.  We let $\route[i]$ indicate the $i$th contact in the sequence and $|\route|$ the length of $\route$ (in the example $\route[0]={S\to A}$ and $|\route|=2$).
In addition, $\mathit{src}(e)$ and $\mathit{tgt}(e)$ indicate the source and target of contact $e$ respectively.

Based on $\Prob$, a SDP for a partial route $\route\in\Rl_n(\ts)$ can be computed as follows.
\begin{align*}
  \makebox[2.3em][l]{$\SDPCGR(\route,\ts) = {}$} & \\ 
  &
  \left(\prod_{k=0}^{|\route|-1} (1 - p_f(\route[k],\ts) \right) \cdot \Prob_{\mathit{tgt}(\route[|\route|-1])}({\ts} + \varsigma(\route[|\route|-1],\ts)) \\[1ex]
  + \ &
  \sum_{k=0}^{|\route|-1} \left( \prod_{i=0}^{k-1} (1 - p_f(\route[i],\ts)) \right) \cdot p_f(\route[k],\ts)\\[-2ex]
  &
  \phantom{\sum_{k=0}^{|\route|-1} \left( \prod_{i=0}^{k-1} (1 - p_f(\route[i],\ts)) \right)}{}\cdot \Prob_{\mathit{src}(\route[k])}({\ts} + \fdd(\route[k],\ts))
\end{align*}
The first summand of the equation corresponds to the successful transmission of the message through all hops in $\route$.  This probability is estimated as the product of the probability of successfully transmitting in each contact --the probability of success in the $i$th hop is $(1 - p_f(\route[k]))$-- times the likelihood (according to \RU) that the message is succesfully transmitted to destination from the last node of the partial route $\route$ (i.e. $\Prob_{\mathit{tgt}(\route[|\route|-1])}({\ts} + \varsigma)$). Notice that this last probability should be considered at the moment that the message is available in the node, which can only be after the transmission delay $\varsigma(\route[|\route|-1],\ts)$.
%
The second summand estimates the probability of successfully transmitting the message given that some hop in $\route$ failed to transmit at time slot $\ts$.  The $k$th summand here corresponds to the likelihood of successfully transmitting given that the hop $k$ is the first to fail.  This is calculated as the product of the probability of succsesfully transmitting in the first $k-1$ hops (i.e. $\left( \prod_{i=0}^{k-1} (1 - p_f(\route[i],\ts)) \right)$~), times the probability of failing in the $k$th hop ($p_f(\route[k],\ts)$), times the likelihood (according to \RU) that the message is succesfully transmitted to destination from the node that failed to transmit in the $k$th hop (i.e. $\Prob_{\mathit{src}(\route[k])}({\ts} + \fdd(\route[k],\ts))$).
Notice this last probability should be considered at the moment that such node detects that the communication has failed, i.e. at ${\ts} + \fdd(\route[k],\ts)$.

The resulting metric $\SDPCGR$ indicates the delivery probability of each route in $Rl_n(t_s)$ computed by CGR, which can be used to decide on a reliable proximate node to forward the bundle with a simple modification to existing implementations.
It is worth noting that \RU might have explored more routes (potentially more reliable) than those in $Rl_n(t_s)$, the construction of which is guided by best delivery time as per CGR's internal Dijkstra searches.
Nevertheless, in Section~\ref{sec:analysis} we show that the \RU-based SDP metric outperforms baseline CGR and approximates the theoretical outcome of \RU and \IRU in random and realistic application scenarios.

\section{Result Analysis}
\label{sec:analysis}

In this section, we propose a benchmark ecosystem to evaluate the proposed routing schemes for DTNs under uncertain contact plans, and use it to analyze the network performance when applying \RU, \IRU and \CGRU. 

\subsection{Benchmark}

A benchmark for DTNs under uncertain contact plans needs to comprise all possible routing solutions that can be considered for such scenarios.
In particular, CGR, sought for fully scheduled DTNs and S\&W, sought for fully unpredictable DTNs sit at the edges of the uncertain DTNs classification.
Other intermediate schemes present in the literature are also considered.
Table~\ref{tab:routing} summarizes and compares the routing schemes present in the benchmark.
We briefly recapitulate them as follows.

\begin{itemize}
\item Upper bound reference:
    \begin{itemize}
    \item \textbf{CGR-FA}: CGR-FA is an oracle-based fault-aware (FA) scheme.
    It leverages the same single-copy implementation than CGR, but uses a contact plan where contacts that will fail are removed.
    By being able to know where and when faults will occur, CGR-FA is used as a theoretical upper bound providing the best achievable performance (delivery ratio and energy consumption).
    \end{itemize}
\item Single-copy, certain contact plan:
    \begin{itemize}
    \item \textbf{CGR}: Current implementation of CGR~\cite{FRAIRE2021102884} in ION v3.5.0~\cite{Burleigh2007} which forwards a bundle using the first contact of the route which has the \textit{best delivery time} among all to the given destination. 
    CGR assumes all contacts in the contact plan will occur as planned.
    \end{itemize}
\item Single-copy, uncertain contact plan:
    \begin{itemize}
    \item \textbf{CGR-HOP}: A variant of CGR which forwards a bundle on the first contact or hop of the route which has the \textit{least hop count} among all to the given destination.
    As discussed in~\cite{Madoery:Congestion}, reducing the hops increases the delivery probability in uncertain contact plans, at the expense of delivery delay.
    \item \textbf{\CGRU}: The \RU-enhanced CGR formulation presented in Section~\ref{sec:rucgr} that enables a straightforward implementation to leverage \RU model features in DTN nodes based on ION protocol stack.
    \end{itemize}
\item Multi-copy, uncertain contact plan:
    \begin{itemize}
    \item \textbf{\RU}: Static routing rules are sent to each node in the network.
    These routes are computed using the \RU model in Algorithm~\ref{Alg:ru}.
    To determine the current state and decide on the subsequent action, nodes have access to a global view of the copy distribution on the network, which is not necessarily feasible in reality.
    The benchmark considers \RU with 1, 2, 3 and 4 copies.
    \item \textbf{\IRU}: Static routing rules are sent to each node in the network by means of the $\LTr$ table. 
    The table comprises a set of specific routing decisions, based on \RU model computed for each node, destination and number of copies. 
    For each bundle, nodes decide routing based on the number of local copies. The benchmark considers \IRU with 1, 2, 3 and 4 copies.
    \item \textbf{CGR-2CP}: Another variant of CGR where two-copies (2CP) are generated at the source~\cite{Madoery:Congestion}.
    Copies are forwarded via both the best delivery time and the least hop count routes, when different.
    CGR-2CP provides equal or better delivery ratio than CGR-HOP with improved delivery delay.
    \end{itemize}
\item Multi-copy, no contact plan knowledge:
    \begin{itemize}
    \item \textbf{S\&W}: Spray-and-wait routing provides similar performance metrics than flooding with less overhead~\cite{Spyropoulos05sprayandwait}.
    The traffic source spreads a limited number of copies to the first contacted neighbors and then wait until one of those copies reaches the destination. We evaluate S\&W with 2, 3 and 4 copies.
    \end{itemize}
\end{itemize}

\newcommand*\rot{\rotatebox{90}}
\newcommand*\OK{\ding{51}}

\begin{table} \centering
\begin{tabular}{@{} cr*{6}c }
  & & \multicolumn{6}{c}{} \\[2ex]
\rowcolor{blue!30} \cellcolor{white}
   & & \rot{Contact plan} & \rot{\shortstack[l]{Encoded\\probability}} & \rot{\shortstack[l]{Encoded failures\:\: \\ (oracle)}} & \rot{\shortstack[l]{Implementable\\(local view)}} 
   & \rot{Copies} & \rot{\shortstack[l]{Main\\optimization\\metric}} \\
    \cmidrule{2-8}
\rowcolor{black!15} \cellcolor{white}
   & CGR-FA     & Yes & Yes & Yes & No & 1 & Delivery   \\
   & RUCoP      & Yes & Yes & No & No  & 1-4 & Delivery \\
\rowcolor{black!15} \cellcolor{white}
   & L-RUCoP    & Yes & Yes & No & Yes & 1-4 & Delivery \\
   & CGR-UCoP   & Yes & Yes & No & Yes & 1   & Delivery \\
\rowcolor{black!15} \cellcolor{white}
   & CGR        & Yes & No & No & Yes & 1    & Delay    \\
   & CGR-HOP    & Yes & No & No & Yes & 1    & Delivery \\
\rowcolor{black!15} \cellcolor{white}
   & CGR-2CP    & Yes & No & No & Yes & 2   & Delivery \& Delay \\
  \rot{\rlap{~Routing Algorithms}}
   & S\&W       & No  & No & No & Yes & 2-4 & Delivery \& Delay \\
    \cmidrule[1pt]{2-8}
    \end{tabular}
    \caption{Routing Schemes in the Benchmark}
    \label{tab:routing}
\end{table}

For each routing scheme, the benchmark considers and evaluates the following routing metrics.
\begin{itemize}
\item \textbf{Delivery Ratio:} number of bundles successfully delivered over number of bundles generated, excluding copies.
This is the main metric of the benchmark.
\item \textbf{Delivery Delay:} mean delay per bundle successfully delivered to the destination.
Non delivered bundles are not considered in the metric; thus, this metric should be considered after the delivery ratio.
\item \textbf{Energy Efficiency:} number of bundles successfully delivered over the total number of transmissions in the network.
Also observed after the delivery ratio, as good efficiency might come at the expense of poor delivery.
\end{itemize}

We analyze the results obtained from two benchmark scenarios: random networks and ring-road networks (RRN).
The former renders a highly connected network with several route paths, while the latter comprises two realistic and simple topologies where satellites can contact ground spots (RRN-A and RRN-B).
In all cases, bundles sizes are set small enough to avoid congestion biases.
Also, channels are configured as error-free (i.e., no packet drop) in order to focus the analysis only on the uncertainty phenomena.

\begin{itemize}
\item \textbf{Random Networks:} Composed of 10 random topologies with 8 nodes and a duration of 100 seconds. 
Time is fragmented in episodes of 10 seconds.
In each episode, the connectivity between nodes (i.e., presence of contacts) is decided based on a contact density parameter of 0.2, similar to~\cite{Madoery2018}.
An all-to-all traffic pattern is assumed.
Each routing algorithm is simulated 100 times on each of the 10 networks and then averaged.
\item \textbf{RRN-A with ISL:} The RRN-A is based on a realistic low-Earth orbit Walker constellation of 16 satellites proposed and described in~\cite{Fraire2017-Hindawi}.
Satellites act as data-mules by receiving data from 22 isolated ground terminals, store it and deliver it to a ground station placed in Argentina.
This is an all-to-one traffic pattern.
In this case, satellites are equipped with Inter-Satellite Links (ISLs) implying contacts are also possible in-orbit~\cite{Fraire2017-Hindawi}.
Routes can thus involve multiple hops between satellites and ground terminals.
The scenario is propagated for 24 hours and sliced into 1440 time slots, each of 60 seconds.
Within a time slot, a contact is considered feasible if a communication opportunity of more than 30 seconds exists.
This corresponds to a fine-grained model.
\item \textbf{RRN-B without ISL:} A different Walker constellation topology of 12 satellites on polar orbits where no close-distance crossing is present.
Not having ISL implies the routes to a target ground spot destination use at most one data-mule satellite.
In this case, the routing decision is taken by a centralized mission control for data flowing from Internet to the isolated terminals. 
This is a one-to-one traffic pattern where routing implies deciding which ground station will be used to upload the data to which satellite.
Two ground stations are configured as gateways in Antarctica and Svalbard.
This scenario considers a coarse-grain model: time slots are defined in such a way that contacts start and terminate within the time slot duration.
\end{itemize}


\begin{figure*}
	\centering
	\includegraphics[width=\textwidth]{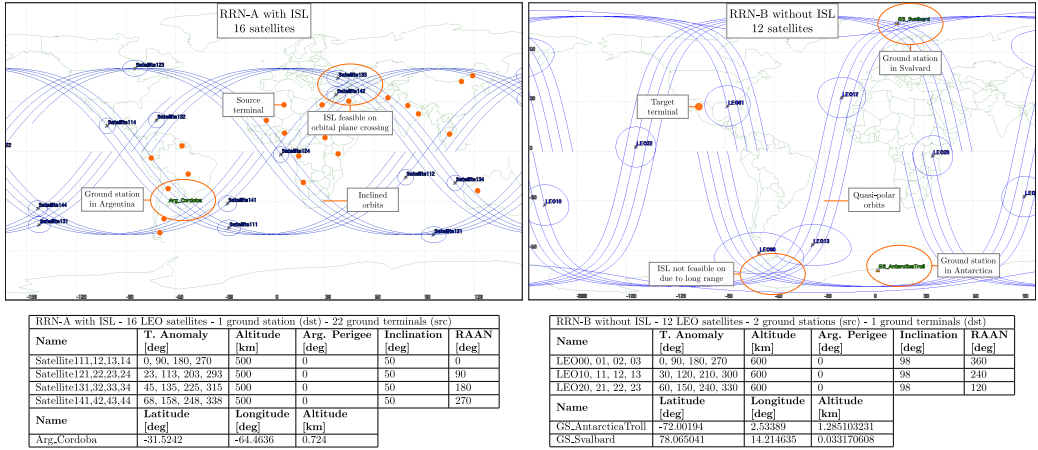}
	\caption{RRN satellite constellation topologies, parameters and orbital tracks. On the left, RRN-A with ISL shows the 22 ground nodes (sources of data) as well as the target ground station in Argentina (many-to-one traffic). On the right, RRN-B without ISL shows the two ground station that can be used as gateways to reach a single target spot (one-to-one traffic).}
	\label{fig:walker}
\end{figure*}

It is worth mentioning that orbital paths\footnote{STK scenarios, visualizations, orbital parameters and ground locations as well as resulting contact plans for the proposed benchmark are publicly available at \url{https://sites.google.com/unc.edu.ar/dtsn-scenarios}} are calculated from STK~\cite{stk} and encoded into contact plans with contact plan designer~\cite{cpd-designer}.
For the sake of simplicity, contact failure probabilities $p_f$ are configured homogeneously in all links, ranging between [0,1].
Indeed, $p_f$ s is the independent variable in the benchmark.
As a result, it is expected that certain contact plan routing provide good metrics when $p_f \approx 1$, while non contact plan based solutions on $p_f \approx 0$.
The hypothesis is that uncertain contact plan approaches outperform both in intermediate values of $p_f$.
By running a large routing simulation campaign using DtnSim~\cite{Fraire:2017:DtnSim}, we are able to determine on which ranges of $p_f$ the hypothesis holds.

\subsection{Results}

\begin{figure*}
    \centering
    \includegraphics[width=\textwidth,trim={3cm 0.8cm 3cm 2cm},clip]{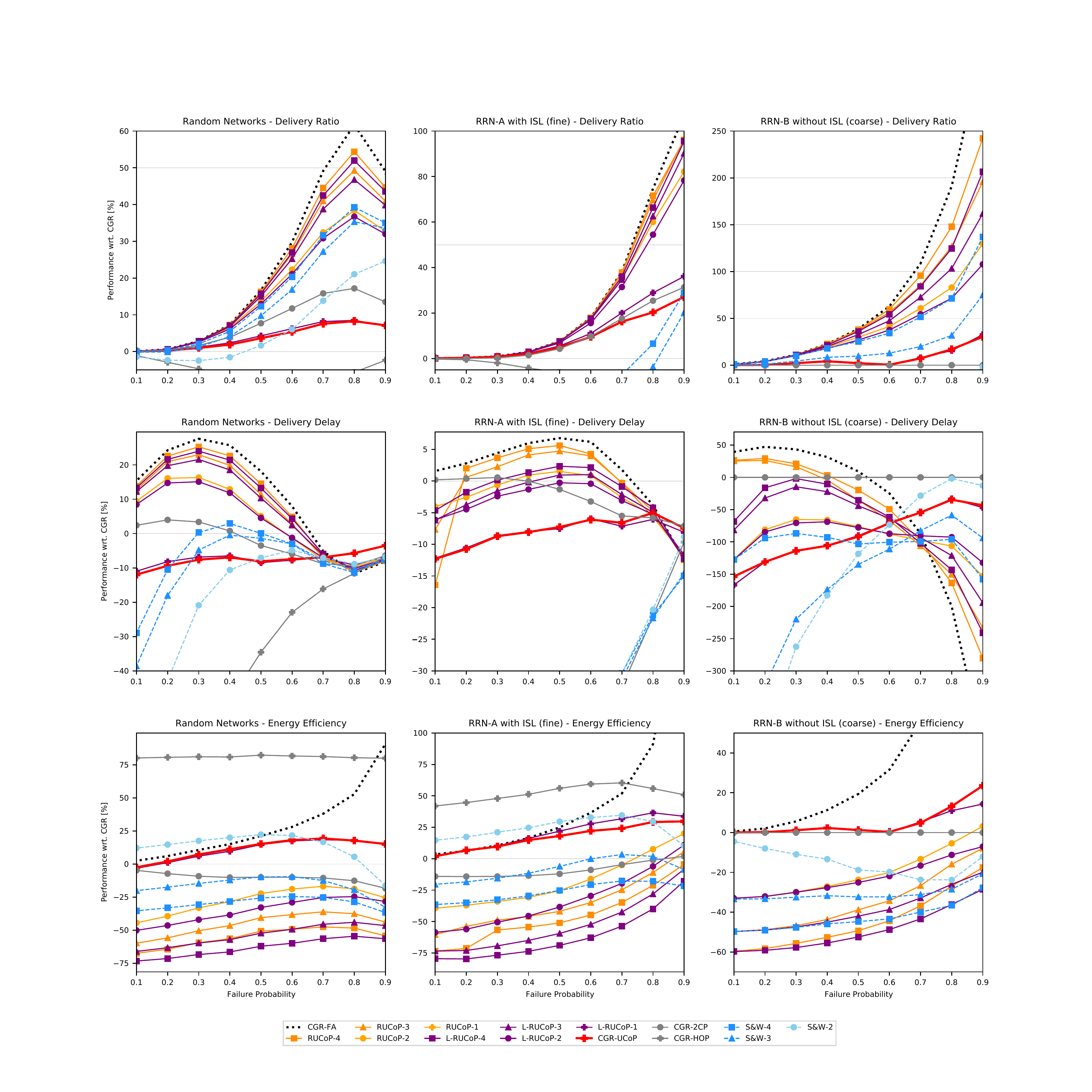}
    \caption{Routing for DTNs under uncertain contact plan benchmark. From left to right, the different scenarios: random networks, RRN-A, and RRN-B. From top to bottom, the different metrics: delivery ratio, delivery delay, energy efficiency. Delivery delay and energy efficiency have to be considered after delivery ratio, as they are computed from delivered bundles only. Curves includes CGR-FA (oracle), \RU (1 to 4 copies), \IRU (1 to 4 copies), \CGRU (adapted CGR), CGR-2CP (two-copies), CGR-HOP (lowest hop count metric), and S\&W (2 to 4 copies).}
    \label{fig:results}
\end{figure*}

The benchmark results\footnote{The \RU implementation in Python3 as well as the scripts used to obtain the results presented in this sections are publicly available at \url{https://bitbucket.org/fraverta/experiments-paper-ieee-tmc-2020}.} are summarized in Fig.~\ref{fig:results}.
To facilitate the comparison with state-of-the-art solutions, metrics are plotted with respect to CGR.
CGR-FA is plotted as maximum theoretical bound in dotted lines.
Because the RRN satellite networks offer simpler (and less) routes (i.e., less hop count) than the random networks, the potential improvement evidenced by CGR-FA in these scenarios is significant towards cases with higher failure probabilities (right hand-side of the curves).

\subsubsection{Delivery Ratio}

When contact failure probabilities are close to $0$, the contact plan occurs as expected (i.e., no uncertainties). 
In this context, and for all studied scenarios, \RU, \IRU and \CGRU provide the same delivery ratio performance than CGR.
Being based on CGR calculations, CGR-2CP and CGR-HOP also provide the same delivery ratio metric.
On the other hand, S\&W algorithms offer limited relative performance in these cases as they have no consideration of the topological knowledge imprinted in the contact plan.

As the probability of failure increases, the delivery ratio diverges for most techniques.
In all scenarios, and for each number of copies, \RU model provides the best delivery ratio results, improving as the number of allowed copies increases.
This improvement becomes more evident for larger $p_f$.
\IRU follows \RU closely, with a delta of performance explained by the fact of solely relaying on (a pessimistic) local node's knowledge.
Also, as expected, S\&W improves the delivery ratio on scenarios with higher uncertainty.
Depending on the number of copies, S\&W schemes can even outperform CGR baseline in particular cases, as already indicated in~\cite{Madoery2018}.
In random networks, S\&W provides good two-copies results, in comparison with CGR-2CP; however, the latter behaves better in simpler networks such as RRN (delivery ratio for S\&W-2 in RRN-A and B is always worst than CGR baseline and thus not plotted).
Nevertheless, \IRU offers the best single-copy implementable routing solution, closely followed by \CGRU, both improving CGR delivery ratio in cases with medium and high failure probabilities.
Moreover, in practical RRN scenarios, \CGRU also provides better performance than S\&W with two copies, and even better than S\&W-3 in RRN with ISL.
Indeed, \IRU with one copy provides the same outcomes than \RU-1, and remarkably, \CGRU (also single-copy) almost always delivers the same performance than both (notice cross markers of \RU and \IRU are behind \CGRU in most of the plots).
This is compelling evidence that the practical applicability of \CGRU can provide great value at minimum implementation costs.
In particular, under high uncertainty, \CGRU outperforms CGR by 9\% in random networks, 22\% in RRN-A with ISL and 25\% in RRN-B without ISL.

\subsubsection{Delivery Delay}

Although not specifically optimized for delivery delay, \RU and \IRU models exhibit a reasonable performance with respect to CGR in this metric, especially in random networks.
This can be explained by the fact that \RU-based models consider all possible paths and can determine the optimal one, which is not always the case of CGR as already discussed in~\cite{FRAIRE201831}.
As $p_f$ increases, the delivery delay of \RU decreases with respect to CGR, but with a much larger deliver ratio. That is, the few bundles that arrive with CGR do so in a shorter time on routes whose contacts do not present failures, while \RU is able to deal with failures and deliver a greater number of bundles, some of which take longer to arrive thereby increasing the average delay value. 
On the implementable side, \CGRU delivery delay performance approaches CGR as the failure probabilities increases.
In realistic RRN scenarios, \CGRU is consistently better than S\&W routing as well as CGR-HOP which honors low hops and potentially higher latency routes (delivery delay for CGR-HOP is the lowest of all schemes not reaching the scale of RRN-A and B plots).
Notably, CGR-2CP offers very similar performances than plain CGR as one of the two copies follows the same lowest delivery delay route than CGR.

\subsubsection{Energy efficiency}

On the energy efficiency side, we care about the transmission effort required to deliver the bundles.
Naturally, single copies schemes offer the least effort, especially CGR-HOP which also minimises the overall hops and thus, transmissions.
On the other hand, multiple copy solutions including \RU-4, \IRU-4 and S\&W-4 demand the largest energy effort, being the latter consistently better, at the expense of a lower delivery ratio.
Remarkably, and being a single copy scheme, \CGRU always offer the same or better energy efficiency than CGR, and is only outperformed by the less performing CGR-HOP and by S\&W-2 in some cases.

To wrap up, \RU model proved to approach the ideal fault-aware case of CGR-FA by leveraging the presented MDP formulation, especially with larger number of copies.
While \RU model can serve as a routing solution with global view, \IRU obtains similar results based on a reduced local view in practical DTNs, and implemented in existing protocol stacks by means of \CGRU.
Indeed, \CGRU has shown that the consideration of the adapted SDP calculation of \RU enables a very appealing performance over the whole failure probability range in DTNs under uncertain contact plan.

\subsection{Discussion}
\label{sec:discussion}

To properly frame the benefits and applicability of \RU and \IRU models and \CGRU algorithm, we discuss some considerations.

\minisection{Multiple Senders:} Although \RU model, as presented in Section~\ref{sec:model}, takes one sender and one destination as arguments, multiple senders can be considered in a single MDP if they seek to reach the same destination.
Indeed, this was already accounted for in the RRN-A case (all-to-one traffic shape), where the same \RU was solved for each of the 22 senders. 
Indeed, a policy was derived for each data flow from a single execution of the MDP.
This can be achieved because the MDP tree for each case is exactly the same except the initial state at $\mathcal{T}_0$.
In general, this approach can be generalized as long as different data flows do not compete for a same limited channel resource (i.e., congestion).

\minisection{Congestion:} In general, congestion is an open research issue in DTN~\cite{silva2015survey}.
In this context, \RU-based models have been sought for and evaluated in scenarios where congestion is not present.
This means that when a route is determined for a bundle, it is assumed that there will be enough capacity to allocate such data transmission (i.e., sizes of the bundles is by far smaller than the contact capacity).
While this can be the case for unsaturated networks, congested networks would need to rely on simulations analysis that validates if the \RU routing assumptions holds. 

\minisection{Scalability:} Table~\ref{tab:scalability} summarizes the scalability metrics of the evaluated scenarios when using \RU.
In particular, the execution time on an Intel i7 processor with 16 GB of RAM running an Ubuntu 19.10 was measured for a Python3 implementation of the \RU routine. 
The explored states and evaluated transitions were listed to observe their increment with larger scenarios and required copies.
Results show that \RU is well suited to solve realistic cases in reasonable time. 
Indeed, less than an hour is required for the more complex case of RRN-A with ISL and four copies of the data.
As already explained, a coarse model of the network offers significant gain in processing time, at the expense of less accurate results.

Compared with the computation time required by \RU, calculating \IRU routing matrix demands a reduced overhead.
The specific processing time for each of the case studies is reported in the \IRU Time[sec] column, in Table~\ref{tab:scalability}.
In particular, the time required for computing \IRU-2 for the RRN-A scenario is the sum of those for \RU-1 and \RU-2 (i.e., $ 258 + 291 = 549$ seconds), plus the cost of building the \IRU routing matrix (37.49 seconds), adding up for a total of $586.49$ seconds. 
As \RU computation can be done in parallel, the time can be significantly reduced.


\begin{table}
\label{tab:scalability}
\caption{Scalability Metrics}
\centering
\begin{tabular}{|l|l|l|l|l|}
\hline
Copies             & 1         & 2         & 3       & 4        \\ \hline
\multicolumn{5}{|l|}{\textbf{Random Networks}}                  \\ \hline
Time {[}sec{]}     & 2         & 6         & 107     & 2416     \\ \hline
States             & 74        & 318       & 1056    & 2915     \\ \hline
Transitions        & 391       & 9491      & 179797  & 2804864  \\ \hline
{\IRU Time[sec]}   & +0.15     & +0.42     & +0.85  & +1.51     \\ \hline 
\multicolumn{5}{|l|}{\textbf{RRN-A with ISL (fine grain)}}      \\ \hline
Time {[}sec{]}     & 258       & 291       & 657     & 3290     \\ \hline
States             & 6091      & 76428     & 646152  & 4126765  \\ \hline
Transitions        & 6973      & 99742     & 969861  & 7147805  \\ \hline
{\IRU Time[sec]}   & +12.92     & +37.49   & +107.19 & +426.96  \\ \hline
\multicolumn{5}{|l|}{\textbf{RRN-B without ISL (coarse grain)}} \\ \hline
Time {[}sec{]}     & 18        & 21        & 38      & 134      \\ \hline
States             & 898       & 8568      & 49774   & 220745   \\ \hline
Transitions        & 1020      & 11133     & 73566   & 369689   \\ \hline
{\IRU Time [sec]}   & +1.75     & +4.38     & +9.02  & +21.16   \\ \hline
\end{tabular}
\end{table}

\section{Conclusion}
\label{sec:conclusion}

Delay Tolerant Networks (DTNs) classification has biased the research of routing algorithms to fit either fully scheduled or dynamically-learned probabilistic use cases.
In this paper, we have uncovered that routing under uncertain contact planning deserves a different classification. 
Uncertain DTNs have not only applicable relevance but also can serve as a more generic routing approach for many practical DTNs.

A first Markov Decision Process coined \RU was introduced for arbitrary number of copies in uncertain DTNs.
\RU provides a theoretical upper bound for the data delivery ratio when a global vision of the system is possible.
\RU enabled the derivation of \IRU when knowledge is restricted to a local view, and single-copy \CGRU where the outcomes of the MDP model can drive routing decisions of the popular CGR routing algorithm.

To evaluate \RU, \IRU and \CGRU, we have proposed an appealing benchmark comprising random and realistic case studies as well as candidate routing solutions.
Results showed that \RU and \IRU models approach the ideal case as the number of copies increases.
On the other hand, single-copy \CGRU has also provided outstanding results under uncertain contact plans, outperforming both CGR (scheduled routing) by up to 25\% in realistic satellite DTNs with uncertain links.

Future work involves the comparison with the simulation results reported in~\cite{d2020sampling} as well as further research on multi-objective optimizations comprising delivery delay and route reliability for \CGRU, which will be implemented and proposed for NASA's ION protocol stack.
Succeeding in such endeavor would settle \CGRU as the de-facto routing scheme for DTNs with uncertain contact plans.

\section*{Acknowledgement}This research has received support from the ERC Advanced Grant 695614 (POWVER), the DFG grant 389792660, as part of TRR 248 (\url{https://perspicuous-computing.science}), the Agencia I$+$D$+$i grant PICT-2017-3894 (RAFTSys), PICT-2017-1335, and the SeCyT-UNC grant 33620180100354CB (ARES).
Part of this work has been developed while Dr. Juan Fraire was visiting Politecnico di Torino.

\bibliographystyle{elsarticle-num} 
\bibliography{biblio}

\begin{thebibliography}{10}
\expandafter\ifx\csname url\endcsname\relax
  \def\url#1{\texttt{#1}}\fi
\expandafter\ifx\csname urlprefix\endcsname\relax\def\urlprefix{URL }\fi
\expandafter\ifx\csname href\endcsname\relax
  \def\href#1#2{#2} \def\path#1{#1}\fi

\bibitem{Fall2003}
K.~Fall, \href{http://doi.acm.org/10.1145/863955.863960}{A delay-tolerant
  network architecture for challenged {I}nternets}, in: Proceedings of the 2003
  Conference on Applications, Technologies, Architectures, and Protocols for
  Computer Communications, SIGCOMM '03, ACM, New York, NY, USA, 2003, pp.
  27--34.
\newblock \href {https://doi.org/10.1145/863955.863960}
  {\path{doi:10.1145/863955.863960}}.
\newline\urlprefix\url{http://doi.acm.org/10.1145/863955.863960}

\bibitem{RFC4838}
V.~Cerf, S.~Burleigh, A.~Hooke, L.~Torgerson, R.~Durst, K.~Scott, K.~Fall,
  H.~Weiss, \href{http://www.rfc-editor.org/rfc/rfc4838.txt}{Delay-tolerant
  networking architecture}, RFC 4838, RFC Editor (April 2007).
\newline\urlprefix\url{http://www.rfc-editor.org/rfc/rfc4838.txt}

\bibitem{Burleigh2003}
S.~Burleigh, A.~Hooke, L.~Torgerson, K.~Fall, V.~Cerf, B.~Durst, K.~Scott,
  H.~Weiss, \href{http://dx.doi.org/10.1109/MCOM.2003.1204759}{Delay-tolerant
  networking: An approach to interplanetary internet}, Comm. Mag. 41~(6) (2003)
  128--136.
\newblock \href {https://doi.org/10.1109/MCOM.2003.1204759}
  {\path{doi:10.1109/MCOM.2003.1204759}}.
\newline\urlprefix\url{http://dx.doi.org/10.1109/MCOM.2003.1204759}

\bibitem{Caini2011}
C.~Caini, H.~Cruickshank, S.~Farrell, M.~Marchese, Delay- and
  disruption-tolerant networking ({DTN}): An alternative solution for future
  satellite networking applications, Proceedings of the IEEE 99~(11) (2011)
  1980--1997.
\newblock \href {https://doi.org/10.1109/JPROC.2011.2158378}
  {\path{doi:10.1109/JPROC.2011.2158378}}.

\bibitem{gupta2015survey}
L.~Gupta, R.~Jain, G.~Vaszkun, Survey of important issues in {UAV}
  communication networks, IEEE Communications Surveys \& Tutorials 18~(2)
  (2015) 1123--1152.

\bibitem{BENAMAR2014141}
N.~Benamar, K.~D. Singh, M.~Benamar, D.~E. Ouadghiri, J.-M. Bonnin,
  \href{http://www.sciencedirect.com/science/article/pii/S0140366414001212}{Routing
  protocols in vehicular delay tolerant networks: A comprehensive survey},
  Computer Communications 48 (2014) 141 -- 158, opportunistic networks.
\newblock \href {https://doi.org/https://doi.org/10.1016/j.comcom.2014.03.024}
  {\path{doi:https://doi.org/10.1016/j.comcom.2014.03.024}}.
\newline\urlprefix\url{http://www.sciencedirect.com/science/article/pii/S0140366414001212}

\bibitem{7876231}
J.~{Hom}, L.~{Good}, {Shuhui Yang}, A survey of social-based routing protocols
  in delay tolerant networks, in: 2017 International Conference on Computing,
  Networking and Communications (ICNC), 2017, pp. 788--792.
\newblock \href {https://doi.org/10.1109/ICCNC.2017.7876231}
  {\path{doi:10.1109/ICCNC.2017.7876231}}.

\bibitem{7921980}
F.~Z. {Benhamida}, A.~{Bouabdellah}, Y.~{Challal}, Using delay tolerant network
  for the {I}nternet of {T}hings: Opportunities and challenges, in: 2017 8th
  International Conference on Information and Communication Systems (ICICS),
  2017, pp. 252--257.
\newblock \href {https://doi.org/10.1109/IACS.2017.7921980}
  {\path{doi:10.1109/IACS.2017.7921980}}.

\bibitem{Partan2007}
J.~Partan, J.~Kurose, B.~N. Levine,
  \href{http://doi.acm.org/10.1145/1347364.1347372}{A survey of practical
  issues in underwater networks}, SIGMOBILE Mob. Comput. Commun. Rev. 11~(4)
  (2007) 23--33.
\newblock \href {https://doi.org/10.1145/1347364.1347372}
  {\path{doi:10.1145/1347364.1347372}}.
\newline\urlprefix\url{http://doi.acm.org/10.1145/1347364.1347372}

\bibitem{RFC5050}
K.~Scott, S.~Burleigh, \href{http://www.rfc-editor.org/rfc/rfc5050.txt}{Bundle
  protocol specification}, RFC 5050, RFC Editor (November 2007).
\newline\urlprefix\url{http://www.rfc-editor.org/rfc/rfc5050.txt}

\bibitem{pottner2011performance}
W.-B. P{\"o}ttner, J.~Morgenroth, S.~Schildt, L.~Wolf, Performance comparison
  of {DTN} bundle protocol implementations, in: Proceedings of the 6th ACM
  workshop on Challenged networks, ACM, 2011, pp. 61--64.

\bibitem{Fraire2015}
J.~A. Fraire, J.~M. Finochietto, Design challenges in contact plans for
  disruption-tolerant satellite networks, Communications Magazine, IEEE 53~(5)
  (2015) 163--169.
\newblock \href {https://doi.org/10.1109/MCOM.2015.7105656}
  {\path{doi:10.1109/MCOM.2015.7105656}}.

\bibitem{Fraire2016Traffic}
J.~A. Fraire, P.~G. Madoery, J.~M. Finochietto, Traffic-aware contact plan
  design for disruption-tolerant space sensor networks, Ad Hoc Networks 47
  (2016) 41 -- 52.
\newblock \href {https://doi.org/http://dx.doi.org/10.1016/j.adhoc.2016.04.007}
  {\path{doi:http://dx.doi.org/10.1016/j.adhoc.2016.04.007}}.

\bibitem{Fraire2015Routing}
J.~A. Fraire, J.~Finochietto, Routing-aware fair contact plan design for
  predictable delay tolerant networks, Ad Hoc Networks 25 (2015) 303 -- 313.
\newblock \href {https://doi.org/https://doi.org/10.1016/j.adhoc.2014.07.006}
  {\path{doi:https://doi.org/10.1016/j.adhoc.2014.07.006}}.

\bibitem{Fraire2014Fair}
J.~A. Fraire, P.~G. Madoery, J.~M. Finochietto, On the design and analysis of
  fair contact plans in predictable delay-tolerant networks, Sensors Journal,
  IEEE 14~(11) (2014) 3874--3882.
\newblock \href {https://doi.org/10.1109/JSEN.2014.2348917}
  {\path{doi:10.1109/JSEN.2014.2348917}}.

\bibitem{carosino2018integrating}
M.~Carosino, J.~A. Fraire, J.~A. Ritcey, Integrating scheduled {DTN}s and
  {TDMA}-based {MAC} sublayers: Preliminary results, in: 2018 6th IEEE
  International Conference on Wireless for Space and Extreme Environments
  (WiSEE), IEEE, 2018, pp. 141--146.

\bibitem{FRAIRE2021102884}
J.~A. Fraire, O.~{De Jonckère}, S.~C. Burleigh,
  \href{http://www.sciencedirect.com/science/article/pii/S1084804520303489}{Routing
  in the space internet: A contact graph routing tutorial}, Journal of Network
  and Computer Applications 174 (2021) 102884.
\newblock \href {https://doi.org/https://doi.org/10.1016/j.jnca.2020.102884}
  {\path{doi:https://doi.org/10.1016/j.jnca.2020.102884}}.
\newline\urlprefix\url{http://www.sciencedirect.com/science/article/pii/S1084804520303489}

\bibitem{Araniti2015}
G.~Araniti, N.~Bezirgiannidis, E.~Birrane, I.~Bisio, S.~Burleigh, C.~Caini,
  M.~Feldmann, M.~Marchese, J.~Segui, K.~Suzuki, Contact graph routing in {DTN}
  space networks: overview, enhancements and performance, IEEE Comms. Magazine
  53~(3) (2015) 38--46.
\newblock \href {https://doi.org/10.1109/MCOM.2015.7060480}
  {\path{doi:10.1109/MCOM.2015.7060480}}.

\bibitem{grasic2011evolution}
S.~Grasic, E.~Davies, A.~Lindgren, A.~Doria, The evolution of a {DTN} routing
  protocol-{PRoPHETv2}, in: Proceedings of the 6th ACM workshop on Challenged
  networks, 2011, pp. 27--30.

\bibitem{burgess2006maxprop}
J.~Burgess, B.~Gallagher, D.~D. Jensen, B.~N. Levine, et~al., {MaxProp}:
  Routing for vehicle-based disruption-tolerant networks., in: Infocom, Vol.~6,
  Barcelona, Spain, 2006.

\bibitem{jain2004routing}
S.~Jain, K.~Fall, R.~Patra, Routing in a delay tolerant network, Vol.~34, ACM,
  2004.

\bibitem{Feldmann2017}
M.~Feldmann, F.~Walter, Routing in ring road networks with limited topological
  knowledge, in: 2017 Int. Conf. on Wireless for Space and Extreme Environments
  (WiSEE), 2017, pp. 63--68.

\bibitem{Vahdat00epidemicrouting}
A.~Vahdat, D.~Becker, Epidemic routing for partially-connected ad hoc networks,
  Tech. rep., Duke University, Department of Computer Science (2000).

\bibitem{Spyropoulos05sprayandwait}
T.~Spyropoulos, K.~Psounis, C.~S. Raghavendra, Spray and wait: An efficient
  routing scheme for intermittently connected mobile networks, in: 2005 ACM
  SIGCOMM Workshop on DTN, 2005, pp. 252--259.

\bibitem{spyropoulos2007spray}
T.~Spyropoulos, K.~Psounis, C.~S. Raghavendra, Spray and focus: Efficient
  mobility-assisted routing for heterogeneous and correlated mobility, in:
  Fifth Annual IEEE International Conference on Pervasive Computing and
  Communications Workshops (PerComW'07), IEEE, 2007, pp. 79--85.

\bibitem{8737620}
K.~Sakai, M.-T. Sun, W.-S. Ku, Data-intensive routing in delay-tolerant
  networks, in: IEEE INFOCOM 2019 - IEEE Conference on Computer Communications,
  2019, pp. 2440--2448.
\newblock \href {https://doi.org/10.1109/INFOCOM.2019.8737620}
  {\path{doi:10.1109/INFOCOM.2019.8737620}}.

\bibitem{Burleigh2016}
S.~Burleigh, C.~Caini, J.~Messina, M.~Rodolfi, Toward a unified routing
  framework for {DTN}, in: 2016 IEEE Int. Conf. on Wireless for Space and
  Extreme Environments (WiSEE), 2016, pp. 82--86.

\bibitem{Fraire2017-Hindawi}
J.~A. Fraire, P.~Madoery, S.~Burleigh, M.~Feldmann, J.~Finochietto, A.~Charif,
  N.~Zergainoh, R.~Velazco, Assessing contact graph routing performance and
  reliability in distributed satellite constellations, Hindawi Journal of
  Computer Networks and CommunicationseVol. 2017, Article ID 2830542, 18 pages
  (2017).
\newblock \href {https://doi.org/10.1155/2017/2830542}
  {\path{doi:10.1155/2017/2830542}}.

\bibitem{kalaputapu1995modeling}
R.~Kalaputapu, M.~J. Demetsky, Modeling schedule deviations of buses using
  automatic vehicle-location data and artificial neural networks,
  Transportation Research Record (1995) 44--52.

\bibitem{sahai2006fundamental}
A.~Sahai, R.~Tandra, S.~M. Mishra, N.~Hoven, Fundamental design tradeoffs in
  cognitive radio systems, in: Proceedings of the first international workshop
  on Technology and policy for accessing spectrum, ACM, 2006, p.~2.

\bibitem{hwang1981system}
C.~Hwang, F.~A. Tillman, M.~Lee, System-reliability evaluation techniques for
  complex/large systems: A review, IEEE Transactions on Reliability 30~(5)
  (1981) 416--423.

\bibitem{Liang2017}
Q.~{Liang}, E.~{Modiano}, Survivability in time-varying networks, IEEE
  Transactions on Mobile Computing 16~(9) (2017) 2668--2681.
\newblock \href {https://doi.org/10.1109/TMC.2016.2636152}
  {\path{doi:10.1109/TMC.2016.2636152}}.

\bibitem{Li2015}
F.~{Li}, S.~{Chen}, M.~{Huang}, Z.~{Yin}, C.~{Zhang}, Y.~{Wang}, Reliable
  topology design in time-evolving delay-tolerant networks with unreliable
  links, IEEE Transactions on Mobile Computing 14~(6) (2015) 1301--1314.
\newblock \href {https://doi.org/10.1109/TMC.2014.2345392}
  {\path{doi:10.1109/TMC.2014.2345392}}.

\bibitem{Madoery2017rpic}
P.~Madoery, F.~Raverta, J.~Fraire, J.~Finochietto, On the performance analysis
  of disruption tolerant satellite networks under uncertainties, in:
  Proceedings of the 2017 XVII RPIC Workshop, 2017.

\bibitem{Madoery2018}
P.~G. {Madoery}, F.~D. {Raverta}, J.~A. {Fraire}, J.~M. {Finochietto}, Routing
  in space delay tolerant networks under uncertain contact plans, in: 2018 IEEE
  International Conference on Communications (ICC), 2018, pp. 1--6.
\newblock \href {https://doi.org/10.1109/ICC.2018.8422917}
  {\path{doi:10.1109/ICC.2018.8422917}}.

\bibitem{Raverta2018}
F.~D. {Raverta}, R.~{Demasi}, P.~G. {Madoery}, J.~A. {Fraire}, J.~M.
  {Finochietto}, P.~R. {D’Argenio}, A {M}arkov decision process for routing
  in space {DTN}s with uncertain contact plans, in: 2018 6th IEEE International
  Conference on Wireless for Space and Extreme Environments (WiSEE), 2018, pp.
  189--194.
\newblock \href {https://doi.org/10.1109/WiSEE.2018.8637330}
  {\path{doi:10.1109/WiSEE.2018.8637330}}.

\bibitem{BiancoA95}
A.~Bianco, L.~de~Alfaro, \href{https://doi.org/10.1007/3-540-60692-0\_70}{Model
  checking of probabalistic and nondeterministic systems}, in: P.~S.
  Thiagarajan (Ed.), Foundations of Software Technology and Theoretical
  Computer Science, 15th Conference, Bangalore, India, December 18-20, 1995,
  Proceedings, Vol. 1026 of LNCS, Springer, 1995, pp. 499--513.
\newblock \href {https://doi.org/10.1007/3-540-60692-0\_70}
  {\path{doi:10.1007/3-540-60692-0\_70}}.
\newline\urlprefix\url{https://doi.org/10.1007/3-540-60692-0\_70}

\bibitem{BaierK08}
C.~Baier, J.~Katoen, Principles of model checking, {MIT} Press, 2008.

\bibitem{BaierAFK18}
C.~Baier, L.~de~Alfaro, V.~Forejt, M.~Kwiatkowska,
  \href{https://doi.org/10.1007/978-3-319-10575-8\_28}{Model checking
  probabilistic systems}, in: E.~M. Clarke, T.~A. Henzinger, H.~Veith, R.~Bloem
  (Eds.), Handbook of Model Checking, Springer, 2018, pp. 963--999.
\newblock \href {https://doi.org/10.1007/978-3-319-10575-8\_28}
  {\path{doi:10.1007/978-3-319-10575-8\_28}}.
\newline\urlprefix\url{https://doi.org/10.1007/978-3-319-10575-8\_28}

\bibitem{d2020sampling}
P.~R. D’Argenio, J.~A. Fraire, A.~Hartmanns, Sampling distributed schedulers
  for resilient space communication, in: NASA Formal Methods Symposium,
  Springer, 2020, pp. 291--310.

\bibitem{Puterman:1994}
M.~L. Puterman, Markov Decision Processes: Discrete Stochastic Dynamic
  Programming, 1st Edition, John Wiley \& Sons, Inc., New York, NY, USA, 1994.

\bibitem{FilarKoos:1996}
J.~Filar, K.~Vrieze, Competitive Markov Decision Processes, Springer-Verlag,
  Berlin, Heidelberg, 1996.

\bibitem{Kwiatkowska2011}
M.~Kwiatkowska, G.~Norman, D.~Parker, {PRISM} 4.0: Verification of
  probabilistic real-time systems, in: G.~Gopalakrishnan, S.~Qadeer (Eds.),
  Proc. 23rd International Conference on Computer Aided Verification (CAV'11),
  Vol. 6806 of LNCS, Springer, 2011, pp. 585--591.

\bibitem{eddy1996hidden}
S.~R. Eddy, Hidden {M}arkov models, Current opinion in structural biology 6~(3)
  (1996) 361--365.

\bibitem{tcs/CheungLSV06}
L.~Cheung, N.~A. Lynch, R.~Segala, F.~W. Vaandrager,
  \href{https://doi.org/10.1016/j.tcs.2006.07.033}{Switched {PIOA:} parallel
  composition via distributed scheduling}, Theor. Comput. Sci. 365~(1-2) (2006)
  83--108.
\newblock \href {https://doi.org/10.1016/j.tcs.2006.07.033}
  {\path{doi:10.1016/j.tcs.2006.07.033}}.
\newline\urlprefix\url{https://doi.org/10.1016/j.tcs.2006.07.033}

\bibitem{tcs/GiroDF14}
S.~Giro, P.~R. D'Argenio, L.~M.~F. Fioriti,
  \href{https://doi.org/10.1016/j.tcs.2013.07.017}{Distributed probabilistic
  input/output automata: Expressiveness, (un)decidability and algorithms},
  Theor. Comput. Sci. 538 (2014) 84--102.
\newblock \href {https://doi.org/10.1016/j.tcs.2013.07.017}
  {\path{doi:10.1016/j.tcs.2013.07.017}}.
\newline\urlprefix\url{https://doi.org/10.1016/j.tcs.2013.07.017}

\bibitem{Burleigh2007}
S.~Burleigh, Interplanetary overlay network: An implementation of the {DTN}
  bundle protocol, in: 2007 4th IEEE Consumer Communications and Networking
  Conference, 2007, pp. 222--226.
\newblock \href {https://doi.org/10.1109/CCNC.2007.51}
  {\path{doi:10.1109/CCNC.2007.51}}.

\bibitem{Madoery:Congestion}
P.~G. Madoery, J.~A. Fraire, J.~M. Finochietto,
  \href{http://dx.doi.org/10.1002/sat.1210}{Congestion management techniques
  for disruption-tolerant satellite networks}, International Journal of
  Satellite Communications and Networking (2018) n/a--n/aSat.1210.
\newblock \href {https://doi.org/10.1002/sat.1210}
  {\path{doi:10.1002/sat.1210}}.
\newline\urlprefix\url{http://dx.doi.org/10.1002/sat.1210}

\bibitem{stk}
{AGI Systems Tool Kit (STK)}, \url{http://www.agi.com/STK}.

\bibitem{cpd-designer}
J.~A. {Fraire}, Introducing contact plan designer: A planning tool for
  {DTN}-based space-terrestrial networks, in: 2017 6th International Conference
  on Space Mission Challenges for Information Technology (SMC-IT), 2017, pp.
  124--127.
\newblock \href {https://doi.org/10.1109/SMC-IT.2017.28}
  {\path{doi:10.1109/SMC-IT.2017.28}}.

\bibitem{Fraire:2017:DtnSim}
J.~A. Fraire, P.~G. Madoery, F.~Raverta, J.~M. Finochietto, R.~Velazco,
  {DtnSim}: Bridging the gap between simulation and implementation of
  space-terrestrial {DTN}s, in: Space Mission Challenges for Information
  Technology (SMC-IT), 2017 IEEE Int. Conference on, 2017.

\bibitem{FRAIRE201831}
J.~A. Fraire, P.~G. Madoery, A.~Charif, J.~M. Finochietto,
  \href{http://www.sciencedirect.com/science/article/pii/S1570870518304359}{On
  route table computation strategies in delay-tolerant satellite networks}, Ad
  Hoc Networks 80 (2018) 31 -- 40.
\newblock \href {https://doi.org/https://doi.org/10.1016/j.adhoc.2018.07.002}
  {\path{doi:https://doi.org/10.1016/j.adhoc.2018.07.002}}.
\newline\urlprefix\url{http://www.sciencedirect.com/science/article/pii/S1570870518304359}

\bibitem{silva2015survey}
A.~P. Silva, S.~Burleigh, C.~M. Hirata, K.~Obraczka, A survey on congestion
  control for delay and disruption tolerant networks, Ad Hoc Networks 25 (2015)
  480--494.

\end{thebibliography}

\end{document}